%% file: WLTS_current_ArXiv_sub.tex
\documentclass[useAMS,usenatbib]{pasa}

\usepackage{amsfonts}
\usepackage{amssymb}
\usepackage{graphicx}
\usepackage{amsmath}
\usepackage{enumitem}
\usepackage{url}
\setlist[itemize]{leftmargin=*}
\setlist[enumerate]{leftmargin=*}
\include{aasmacros}

\title[Weak Lensing Probability]{Direct Shear Mapping: Prospects for weak lensing studies of individual galaxy-galaxy lensing systems}
\author[C. O. de Burgh-Day, E. N. Taylor, R. L. Webster and A. M. Hopkins]{C. O. de Burgh-Day$^{1,2,3}$\thanks{cdbd@student.unimelb.edu.au$\,$(CDBD); ent@ph.unimelb.edu.au$\,$(ENT); r.webster@unimelb.edu.au$\,$(RLW); ahopkins@aao.gov.au$\,$(AMH)}, E. N. Taylor$^{1,3}$\footnotemark[1], R. L. Webster$^{1,3}$\footnotemark[1] and A. M. Hopkins$^{2,3}$\footnotemark[1]\\
\affil{$^{1}$School of Physics, David Caro Building, The University of Melbourne, Parkville VIC 3010, Australia}
\affil{$^{2}$The Australian Astronomical Observatory, PO Box 915, North Ryde NSW 1670, Australia}
\affil{$^{3}$ARC Centre of Excellence for All-sky Astrophysics (CAASTRO)}}
\jid{PASA}
\doi{10.1017/pas.\the\year.xxx}
\jyear{\the\year}

\usepackage[authoryear]{natbib}
\bibpunct{(}{)}{;}{a}{}{,}
\begin{document}

\begin{abstract}
We have investigated, using both a theoretical and an empirical approach, the frequency of low redshift galaxy-galaxy lensing systems in which the signature of weak lensing might be directly detectable. We find good agreement between these two approaches. In order to make a theoretical estimate of the weak lensing shear, $\gamma$, for each galaxy in a catalogue, we have made an estimate of the asymptotic circular velocity from the stellar mass using three different approaches: from a simulation based relation, from an empirically-derived relation, and using the baryonic Tully-Fisher relation. Using data from the Galaxy and Mass Assembly redshift survey we estimate the frequency of detectable weak lensing at low redshift. We find that to a redshift of $z\sim 0.6$, the probability of a galaxy being weakly lensed by at least $\gamma = 0.02$ is $\sim 0.01$. A scatter in the $M_*-M_h$ relation results in a shift towards higher measured shears for a given population of galaxies. Given this, and the good probability of weak lensing at low redshifts, we have investigated the feasibility of measuring the scatter in the $M_*-M_h$ relation using shear statistics. This is a novel measurement, and is made possible because DSM is able to make individual \itshape direct~\upshape shear measurements, in contrast to traditional weak lensing techniques which can only make statistical measurements. 
We estimate that for a shear measurement error of $\Delta\gamma = 0.02$ (consistent with the sensitivity of DSM), a sample of $\sim$50,000 spatially and spectrally resolved galaxies would allow a measurement of the scatter in the  $M_*-M_h$ relation to be made. While there are no currently existing IFU surveys of this size, there are upcoming surveys which will provide this data (e.g The Hobby-Eberly Telescope Dark Energy Experiment (HETDEX), surveys with Hector, and the Square Kilometre Array (SKA)). 

\end{abstract}


\maketitle

\section{Introduction}
\label{sec:Introduction}
Weak gravitational lensing is a powerful probe of dark matter in the universe (eg. \citeauthor{1993ApJ...404..441K}~\citeyear{1993ApJ...404..441K}). 
Following initial investigations by \citet{2002ApJ...570L..51B} and \citet{2006ApJ...650L..21M}, \citet{2015MNRAS.451.2161D} have developed a new method to measure the weak lensing signal in individual galaxies called Direct Shear Mapping (DSM). The primary scientific application considered by \citet{2015MNRAS.451.2161D} is the measurement of mass and mass distribution in dark matter halos around individual low-redshift galaxies. In particular, since the dark matter halo properties can be measured for individual galaxies, DSM will enable the measurement of the dispersion in the galaxy luminous matter to dark matter ratio, as a function of other galaxy observables. \\

\noindent The possibility of measuring individual galaxy dark matter halo masses through DSM is an exciting prospect, however the measurement itself is challenging, and potentially observationally expensive. We have consequently developed the approach presented here for identifying the most robust candidates for such a measurement. We have also used this approach to investigate the possibility of measuring the scatter in the $M_*-M_h$ relation using shear statistics, and we estimate the size of the statistical sample that would be required to make this measurement.\\

\noindent In this paper the probability of weak lensing shear has been estimated as a function of the redshifts of the source and lensing galaxies, and a catalogue of candidate galaxy pairs is selected from the Galaxy and Mass Assembly Phase 1 Survey (GAMA I) Data Release 2 (DR2) catalogue (\citeauthor{2011MNRAS.413..971D} \citeyear{2011MNRAS.413..971D}, Liske et al. in prep). 
We also find that the distribution of shears in a galaxy sample is influenced by the relationship between stellar mass and halo circular velocity, and the scatter in this relation. With enough shear measurements it may be possible to constrain this relationship, and to measure the scatter.\\

\noindent DSM uses spatially resolved velocity field information for an object to obtain a shear measurement from the velocity map. DSM assumes intrinsic rotational symmetry in the the velocity map, and searches for departures from this symmetry. This requires either radio data cubes or spatially resolved optical spectroscopy. To identify prospective targets, it is desirable to first obtain an estimate of the shear signal present in a galaxy. \\

\noindent While galaxy-galaxy lensing has been used to measure halo masses in the past, those studies stack many galaxy-galaxy pairs statistically, to obtain average halo masses \citep{1996ApJ...466..623B, 1998ApJ...503..531H,  2001ApJ...555..572W,2006MNRAS.368..715M}. In addition, measurements of galaxy halo shapes have been made from stacked galaxy-galaxy weak lensing measurements \citep{2004ApJ...606...67H, 2012A&A...545A..71V}.\\

\noindent To test our target selection algorithm, the weak lensing statistics of a sample of galaxies in the Galaxy and Mass Assembly Data Release 2 (GAMA-DR2) catalogue were investigated, using the stellar mass estimates from  \citet{2011MNRAS.418.1587T}.
The purpose of our lensing frequency algorithm is to estimate of the distribution of shear signals present in a dataset. This algorithm enables novel measurements to be made, and will improve the success rate of any survey intended to measure weak lensing via the DSM method.\\

\noindent The rest of this paper is organised as follows. In Section~\ref{sec:WeakLensing} relevant weak lensing theory is described. In Section~\ref{sec:LensingProb} a theoretical estimate of the probability of weak lensing at low redshift is made, following \citet{2000MNRAS.319..860M}. In Section~\ref{sec:Lensingfreqalg} the inputs, structure and outputs of the lensing frequency algorithm are outlined. In Section~\ref{sec:ApplicationtotheGAMAsurveycatalogue} results of the application of the algorithm to a dataset obtained from GAMA-DR2 are presented, along with an investigation of the possibility of using shear statistics to measure the scatter in the $M_*-M_h$ relation. Conclusions and a summary are presented in Section~\ref{sec:Conclusions}. Throughout the paper we assume a flat Concordance cosmology, with $\Omega_m = 0.3$, $\Omega_\Lambda = 0.7$, $H_0 = 70$km$\,$s$^{-1}\,$Mpc$^{-1}$, and $h_{70} = H_0 / 100 = 0.7$. 

\section{Weak lensing}
\label{sec:WeakLensing}
In this section we will outline the relevant weak lensing theory. Gravitational lensing is the deflection of light from some source on its path to the observer by an intervening mass. The deflection angle is given by 
\begin{equation}
\boldsymbol{\alpha}(\boldsymbol{\xi}) = \int \mathrm{d}^2\boldsymbol{\xi}^\prime\,\frac{4\,G\,\Sigma(\boldsymbol{\xi}^\prime)}{c^2}\,\frac{\boldsymbol{\xi}-\boldsymbol{\xi}^\prime}{|\boldsymbol{\xi}-\boldsymbol{\xi}^\prime|^2},
\end{equation}
where $G$ is the gravitational constant, $\Sigma$ is the projected mass distribution of the deflector, $c$ is the speed of light and $\xi$ is the distance from the deflector (i.e the impact parameter). Since the angle of incidence of the light has been altered, the source will appear to be in a different location to its true position. The true position of the source can be found by solving the lens equation
\begin{equation}
\label{eq:lens_eq}
\boldsymbol{\beta} = \boldsymbol{\theta} - \frac{D_\mathrm{ds}}{D_\mathrm{s}}\boldsymbol{\alpha},
\end{equation}
where $\theta$ is the apparent angular separation of the deflector and source, $\beta$ is the true angular separation of the deflector and source, $D_\mathrm{ds}$ is the angular diameter distance between the deflector and source and $D_\mathrm{s}$ is the angular diameter distance between the observer and source. The angular coordinates $\boldsymbol{\beta}$ and $\boldsymbol{\theta}$ can be related to the corresponding physical coordinates in the source and lens planes as
\begin{align}
\boldsymbol{\eta} &= D_\mathrm{s}\boldsymbol{\beta}\\
\boldsymbol{\xi} &= D_\mathrm{d}\boldsymbol{\theta}\\
\end{align}
In the case of a circularly symmetric lens, and perfect alignment between the observer, lens and source, the lens equation can be solved to obtain the Einstein radius, a characteristic length scale
\begin{equation}
\theta_\mathrm{E}^2 = \frac{4\pi GM(<\theta_\mathrm{E}D_\mathrm{d})}{c^{2}}\frac{D_\mathrm{ds}}{D_\mathrm{s}D_\mathrm{d}},\label{eq:bend_angle}
\end{equation}
where $M(<\theta_\mathrm{E}D_\mathrm{d})$ is the mass enclosed within the Einstein radius, and $D_\mathrm{d}$ is the observer-deflector angular diameter distance.\\

\noindent In the weak lensing regime the light from the source passes well outside the Einstein radius, and the source is singly-imaged. In this case one can assume that so long as the length scale of the source is much less than that of the deflector, then the lensing will be linear. Thus it can be represented by a first-order Taylor expansion, allowing  Equation~(\ref{eq:lens_eq}) to be re-expressed as a linear coordinate mapping between the lensed and un-lensed coordinate systems
\begin{equation}
\boldsymbol{\beta} = \mathbf{A}\,\boldsymbol{\theta}\label{eq:lensmapping},
\end{equation}
where $\mathbf{A}$ is the Jacobian of transformation;
\begin{equation}
\mathbf{A} = \frac{\partial\boldsymbol{\beta}}{\partial\boldsymbol{\theta}} = \left(\begin{array}{cc}
1-\kappa-\gamma_1 & -\gamma_2\\
-\gamma_2 & 1-\kappa+\gamma_1\\
\end{array}\right)\label{eq:A_ij}.
\end{equation}
Here $\kappa$ is the convergence, and $\gamma_1$ and $\gamma_2$ are the two components of the shear vector $\boldsymbol{\gamma} = (\gamma_1,\gamma_2)$. Equation~(\ref{eq:A_ij}) can be inverted to obtain
\begin{align}
\kappa &= 1-(A_{11}+A_{22})/2 \label{eq:kappa}\\
\gamma_1 &= -(A_{11}-A_{22})/2 \label{eq:gamma1}\\
\gamma_2 &= -A_{21} = -A_{12}  \label{eq:gamma2}.
\end{align}
The shear vector $\boldsymbol{\gamma}$ can be rewritten in polar co-ordinates as a function of a shear magnitude, $\gamma$, and an angle, $\phi$ 
\begin{equation}
\gamma_1 = \gamma\cos\phi; \gamma_2 = \gamma\sin\phi,
\end{equation}
where $\phi$ is the angle of the shear vector and $\gamma = \sqrt{\gamma_1^2+\gamma_2^2}$. In this work we are interested in the value of $\gamma$ for a given lens-source system, which is a function of the lens projected mass density and the lens-source angular separation. \\

\noindent We assume a Singular Isothermal Sphere (SIS) lens profile throughout this paper, whose projected surface density has the form
\begin{equation}
\Sigma(\boldsymbol{\xi}) = \frac{\sigma_v^2}{2G}\frac{1}{|\boldsymbol{\xi}|},
\end{equation}
where $\sigma_v$ is the halo velocity dispersion. Although the SIS profile is a primitive lens model, it is sufficient for this work and allows for a simple shear estimation. The Einstein radius for an SIS lens is given by
\begin{equation}
\theta_\mathrm{E}^2 = \frac{4\pi\sigma_v^2}{c^2}\frac{D_\mathrm{ds}}{D_\mathrm{s}}\label{eq:theta_E_SIS},
\end{equation}
and from e.g. \citet{2009MNRAS.396.2257L}, the shear components for an SIS are
\begin{equation}
\gamma_1 = \frac{D_\mathrm{d}\theta_\mathrm{E}}{2|\boldsymbol{\xi}|^3}(\xi_2^2-\xi_1^2);\,\,\gamma_2 = -\frac{D_\mathrm{d}\theta_\mathrm{E}}{|\boldsymbol{\xi}|^3}\xi_1\xi_2,
\end{equation}
so that 
\begin{equation}
\gamma = \frac{D_\mathrm{d}\theta_\mathrm{E}}{2|\boldsymbol{\xi}|}.
\end{equation}
Thus we have an expression for the shear magnitude as a function of lens-source projected separation and lens velocity dispersion, the latter of which can be related to the halo mass.

\section{An estimation of the probability of lensing}
\label{sec:LensingProb}
In this section we describe the process by which a theoretical estimate may be made of the probability of a given source being weakly lensed, as a function of the source redshift. We begin by introducing a similar calculation for strong lensing in analytically solvable cosmologies from \citet{2000MNRAS.319..860M}. We then adapt this work to the weak lensing case, and for a more realistic Concordance cosmology. In the following sections, we will assume two different lens populations: a population of halos housing elliptical galaxies, as in \citet{2000MNRAS.319..860M}, and a Press-Schechter \citep{1974ApJ...187..425P} population of halos. In both cases we assume an SIS halo. We will discuss the general steps for obtaining the expression for the weak lensing optical depth, and will then discuss the elliptical galaxy and Press-Schechter halo cases individually.\\

\noindent Given an estimate for the spatial distribution and mass of lensing galaxies in a given volume, it is possible to make an estimate of the distribution of shears in the volume. This can in turn be used to make an estimate of the probability distribution of weak shears across the sky, assuming the distribution of lensing galaxies is isotropic. \citet{2000MNRAS.319..860M} have used these arguments to make an estimate of the probability of lensing of quasars by elliptical galaxies for three simplified cosmologies. They define the lensing optical depth, $\tau$, as the fraction of the source plane within which the lens equation has multiple solutions. It can be used as an estimator for strong lensing probability. The contribution to the total optical depth by one lensing galaxy is 
\begin{equation}
\tau_{\mathrm{g}} = \frac{\pi\beta_\mathrm{crit}^{2}}{4\pi},
\end{equation}
i.e. the fraction of the sky covered by its lensing cross-section. Here $\beta_\mathrm{crit}$ is the angular distance from the deflector where a background source transitions between being singly or multiply imaged. \\

\noindent In the case of weak lensing, rather than being interested in the region where the source is multiply imaged, we are interested in the region where the source is singly imaged, yet sheared sufficiently such that it is still a measurable effect. This region takes the form of an annulus about the lens (assuming a spherically symmetric lens), the inner bound of which is the Einstein radius, $\theta_\mathrm{E}$, and the outer bound of which is a function of some limiting shear value, $\gamma_\mathrm{lim}$. Thus, $\tau$ is re-written as a function of a new area, $a(\gamma_\mathrm{lim})$, defined as the area covered by this annulus:
\begin{equation}
\tau_{\mathrm{g}}(\gamma_\mathrm{lim}) = \frac{a(\gamma_\mathrm{lim})}{4\pi},
\end{equation}
and
\begin{equation}
a(\gamma_\mathrm{lim}) = \pi \left[\left(\frac{1}{2\gamma_\mathrm{lim}} - 1\right)\theta_\mathrm{E}\right] ^{2},
\end{equation}
and substituting in equation~(\ref{eq:theta_E_SIS}) this becomes
\begin{equation}
\tau_{\mathrm{g}}(\gamma_\mathrm{lim}) = \pi^{2} \left(\frac{1}{2\gamma_\mathrm{lim}} - 1\right)^{2}\left(\frac{\sigma}{c}\right)^{4}\left(\frac{D_\mathrm{ds}}{D_\mathrm{s}}\right)^{2},
\end{equation}
where $\sigma$ is the velocity dispersion of the deflector, $c$ is the speed of light, $D_\mathrm{ds}$ is the deflector-source distance, and $D_\mathrm{s}$ is the observer-source distance.\\ 

\noindent If it is assumed that the population of lensing objects are non-evolving and have uniform volume density at all redshifts (which is reasonable at low redshift), then the differential number of objects at redshift $z$ with velocity dispersion $\sigma$ is 
\begin{equation}
\frac{\mathrm{d}^{2}N_\mathrm{d}}{\mathrm{d}z_\mathrm{d}\mathrm{d}\sigma} = \frac{\mathrm{d}V_{0}}{\mathrm{d}z_\mathrm{d}}\frac{\mathrm{d}n_\mathrm{d}}{\mathrm{d}\sigma}.
\end{equation}
Here $\mathrm{d}V_{0}/\mathrm{d}z$ is the comoving volume element at redshift $z$,
\begin{equation}
\frac{\mathrm{d}V_{0}}{\mathrm{d}z} = \frac{c}{H_{0}}\frac{(1+z)^{2}D_{A}(z)^{2}}{E(z)}\mathrm{d}\Omega,
\end{equation}
where $c$ is the speed of light, $D_A(z)$ is the angular diameter distance at redshift $z$, and 
\begin{equation}
E(z) = \sqrt{(1+z)^{3}\Omega_{\mathrm{M}}+\Omega_{\Lambda}},
\end{equation}
assuming a flat universe.\\

\noindent The optical depth to a redshift $z_\mathrm{s}$ is then obtained by integrating over the optical depths of the entire population of deflectors up to $z_\mathrm{s}$
\begin{equation}
\label{eq:tau_zs}
\tau(z_{\mathrm{s}},\gamma_\mathrm{lim}) = \int^{z_\mathrm{s}}_0\int^{\infty}_0 \frac{\mathrm{d}^{2}N_\mathrm{d}}{\mathrm{d}z_\mathrm{d}\mathrm{d}\sigma} \tau_\mathrm{g}(\gamma_\mathrm{lim}) \mathrm{d}\sigma\mathrm{d}z_\mathrm{d}.
\end{equation}	

\noindent We will now discuss using this relation to estimate the weak lensing optical depth as a function of source redshift for a population of elliptical galaxies, $n_\mathrm{g}$, and a population of dark matter halos, $n_\mathrm{h}$. 

\subsection{A population of halos housing elliptical galaxies}

\citet{2000MNRAS.319..860M} gives the local comoving number density of elliptical galaxies as 
\begin{equation}
\frac{\mathrm{d} n_\mathrm{g}}{\mathrm{d} \sigma_{\parallel}} = \frac{\delta n_{*}}{\sigma_{*}} \left(\frac{\sigma_{\parallel}}{\sigma_{*}}\right)^{\delta(1+\alpha)-1}\exp\left[-\left(\frac{\sigma_{\parallel}}{\sigma_{*}}\right)^\delta\right],
\end{equation}
where $\sigma_{\parallel}$ is the observed line-of-sight velocity dispersion, $\alpha=-1.07\pm0.05$, $n_{*} = (0.0019\pm0.003)h_{70}^{3}\,$Mpc$^{-3}$, $\sigma_{*} = 225\pm 20\,$km$\,$s$^{-1}$ and $\delta = 3.7\pm 1$. The values of $\alpha$ and $n_{*}$ are drawn from \citet{1988MNRAS.232..431E}, and the values of $\sigma_{*}$ and $\delta$ from \citet{1982ApJ...256..346D}. Substituting this into equation~(\ref{eq:tau_zs}), the inner integral can be solved analytically, to give 
\begin{equation}
\label{equation:tau}
\tau(z_\mathrm{s},\gamma_\mathrm{lim}) = C_\mathrm{g}
\int^{z_\mathrm{s}}_{0} \frac{(1+z_\mathrm{g})^{2}}{E(z_\mathrm{g})} \left(\frac{D_\mathrm{d}D_\mathrm{ds}}{D_\mathrm{s}}\right)^{2}\mathrm{d}z_\mathrm{g},
\end{equation}
where 
\begin{equation}
C_\mathrm{g} = \pi^{2}n_{*}\left(\frac{\sigma_\parallel}{c}\right)^{4} \left(\frac{1}{2\gamma_\mathrm{lim}}-1\right)^2 \Gamma(1+\alpha+\frac{4}{\delta}).
\end{equation}
For a given limiting shear value, $\gamma_\mathrm{lim}$, to obtain the probability of measuring a shear of \itshape{at least} \upshape the limiting value, one must solve equation~(\ref{equation:tau}) numerically as a function of the source redshift, $z_\mathrm{s}$. 

\noindent The dashed grey line in Figure~\ref{fig:Probability_of_lensing_afo_zs} shows the lensing optical depth as a function of source redshift for a limiting shear of $\gamma_\mathrm{lim} = 0.02$ based on a population of elliptical galaxies. The coloured lines in Figure~\ref{fig:Probability_of_lensing_afo_zs} are discussed in the next section. 

\subsection{A Press-Schechter halo population}

The Press-Schechter mass function is given by 
\begin{equation}
\frac{\mathrm{d}n}{\mathrm{d}M} = \sqrt{\frac{2}{\pi}}\frac{\rho_m}{M}\times-\frac{\mathrm{d}\ln\sigma(M)}{\mathrm{d}M}\frac{\delta_{crit}}{\sigma(M)}\exp^{-\frac{\delta^2_{crit}}{2\sigma^2(M)}}
\end{equation}
and gives the co-moving number density of dark matter halos as a function of halo mass \citep{1974ApJ...187..425P}. \\

\noindent Substituting this into equation~(\ref{eq:tau_zs}), 
\begin{equation}
\label{eq:tau_zs_halo}
\tau(z_{\mathrm{s}},\gamma_\mathrm{lim}) = \int^{z_\mathrm{s}}_0\int^{\infty}_{M_\mathrm{min}} \frac{\mathrm{d}^{2}n_\mathrm{d}}{\mathrm{d}z_\mathrm{d}\mathrm{d}M} \tau_\mathrm{d}(\gamma_\mathrm{lim}) \mathrm{d}M\mathrm{d}z_\mathrm{d},
\end{equation}
where $M_\mathrm{min}$ is the mass of the smallest halo capable of containing a galaxy, and 
\begin{equation}
\tau_{\mathrm{g}}(\gamma_\mathrm{lim}) = \pi^{2} \eta \left(\frac{1}{2\gamma_\mathrm{lim}} - 1\right)^{2}\left(\frac{(M\,h_{70}^{-1})^{0.316}}{c}\right)^{4}\left(\frac{D_\mathrm{ds}}{D_\mathrm{s}}\right)^{2},
\end{equation}
where $\eta = 6.14656\times 10^{-7}$ and we have used the relation 
\begin{equation}
V_\mathrm{c} = 2.8 \times 10^{-2} \left(M\,h_{70}^{-1}\right) ^{0.316},
\end{equation}
from \citet{2011ApJ...740..102K}. $V_\mathrm{c}$ is the halo asymptotic circular velocity, and we have assumed $V_\mathrm{c} \simeq \sigma$. 
The lower bound of the integral over $M$ in equation~(\ref{eq:tau_zs_halo}) has been truncated at $M_\mathrm{min}$ because we are only interested in halos large enough to contain at least one galaxy. The value of $M_\mathrm{min}$ at $z\simeq0$ is usually taken to be $\log_{10}(M_\mathrm{min}/M_\odot) \simeq 10$ \citep{2001PhR...349..125B}.\\

\noindent In this case, the integral over mass is not analytically solvable, however the two integrals are still separable, giving the following expression:
\begin{equation}
\begin{split}
\label{eq:tau_zs_halo_2}
\tau(z_{\mathrm{s}},\gamma_\mathrm{lim}) & = C_h \int^{\infty}_{M_\mathrm{min}} \frac{\mathrm{d}n_h}{\mathrm{d}M} \left(M\,h_{70}^{-1}\right) ^{1.264} \mathrm{d}M\\
& \times  \int^{z_\mathrm{s}}_0 \frac{(1+z_h)^{2}}{E(z_h)} \left(\frac{D_\mathrm{d}D_\mathrm{ds}}{D_\mathrm{s}}\right)^{2}\mathrm{d}z_h, 
\end{split}
\end{equation}
where 
\begin{equation}
C_h = \pi^2\eta^2 \left(\frac{1}{2\gamma_\mathrm{lim}} - 1\right)^{2}\left(\frac{\sigma^*}{c}\right)^{4},
\end{equation}
and $\sigma^* \simeq 321$ km$\,$s$^{-1}$ is a halo characteristic velocity dispersion, corresponding to a characteristic halo mass $M^*$ defined as where $\sigma(M^*) = \delta_c(z)$, where $\delta_c(z)$ is the critical density and $\delta_c(z=0) \simeq 1.686$. At $z=0$, $\log_{10}(M^*/M_\odot) \simeq 13$ \citep{2001PhR...349..125B}.\\

\begin{figure}
\begin{center}
\includegraphics[width=8.5cm]{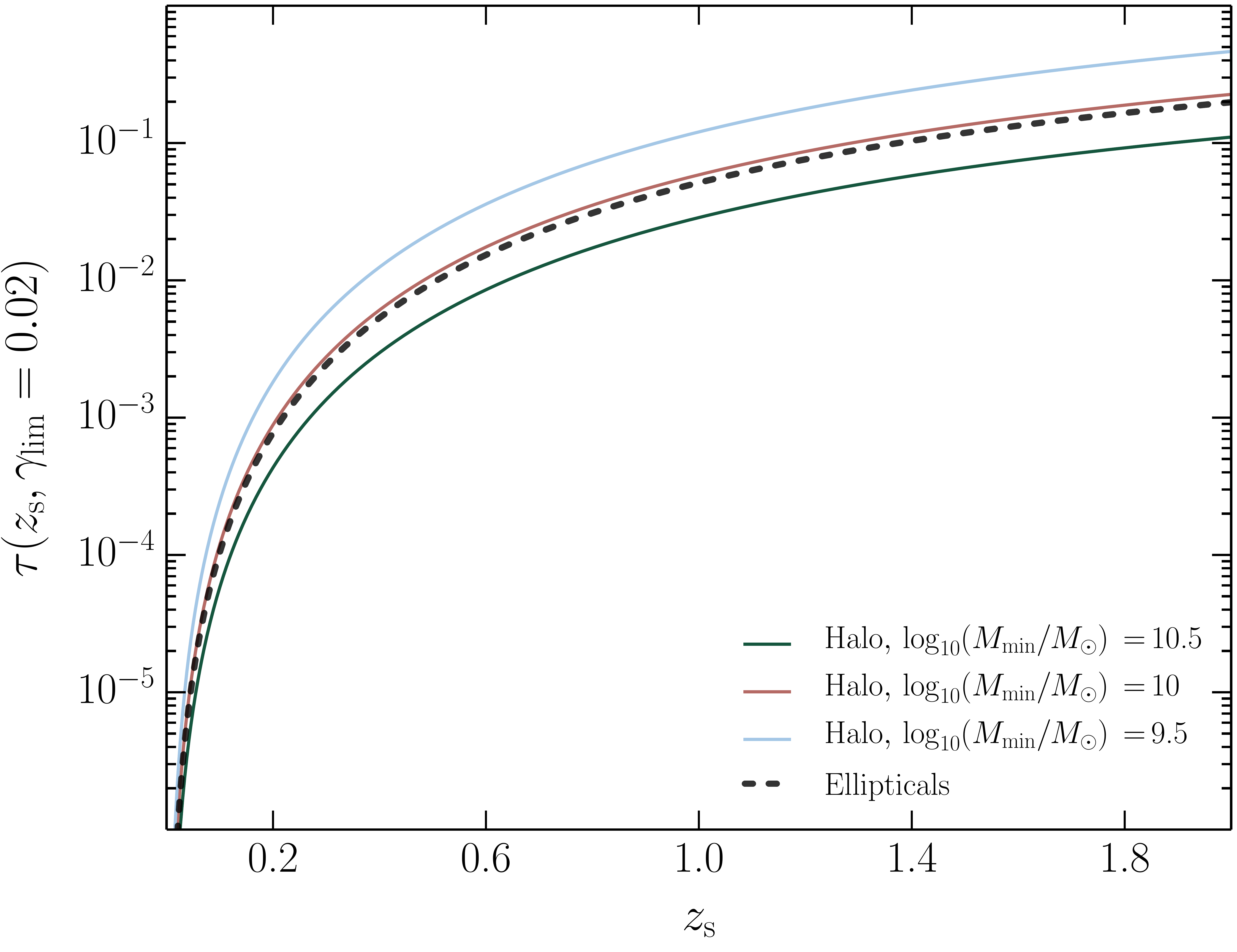}
\caption{The lensing optical depth as a function of source redshift and a limiting shear of $\gamma_\mathrm{lim} = 0.02$. The coloured solid lines show the optical depths obtained when using a population of lenses drawn from a Press-Schechter halo mass function. The three lines show the effect of different minimum halo masses. The minimum halo mass is usually taken to be $10^{10}$M$_\odot$ (e.g. in the Millennium simulation). The dashed grey line shows the optical depth obtained when using a population of lenses drawn from an elliptical galaxy population. It is reassuring to see that the Press-Schechter curve which best matches the dashed curve is that which uses the commonly used minimum halo mass of $10^{10}$M$_\odot$.}
\label{fig:Probability_of_lensing_afo_zs}
\end{center}
\end{figure}

\noindent To obtain the probability of a shear of at least $\gamma_\mathrm{lim}$ from halos, as a function of source redshift, the mass and redshift integrals in equation~(\ref{eq:tau_zs_halo_2}) must be solved numerically. In order to illustrate the effect of varying the minimum halo mass, the mass integral in equation~(\ref{eq:tau_zs_halo_2}) has been solved for three values of the minimum halo mass: $\log_{10}( M_\mathrm{min} / M_\odot )= 9.5,10,$ and $10.5$. The results of this are shown by the three coloured curves in Figure~\ref{fig:Probability_of_lensing_afo_zs}. As expected, the probability curve which best matches that of an elliptical population is for the case $\log_{10}( M_\mathrm{min} / M_\odot ) = 10$. Not surprisingly, the higher the minimum halo mass the lower the probability of lensing, since the majority of halos are of lower mass.\\

\noindent From Figure~\ref{fig:Probability_of_lensing_afo_zs}, roughly 1 in 1,000 sources at $z \sim 0.2$ will be sheared by at least $\gamma = 0.02$. In the calculation of halo lensing probability, four major assumptions have been made:
\begin{itemize}
\item The lensing cross-sections of each galaxy in the population do not overlap. This assumption breaks down if: 1) The redshift is high, since the number of lenses contributing to the optical depth increases with increasing source redshift. 2) $M_\mathrm{min}$ is small, since this leads to more lenses and an increasing cross-section. 3) $\gamma_\mathrm{lim}$ is small, since the cross-section of each lens is larger for a smaller limiting shear. 
\item Both assumed halo populations do not evolve with redshift. This assumption is only correct over small redshift ranges. The elliptical and halo populations used in Figure~\ref{fig:Probability_of_lensing_afo_zs} are for $z = 0$. 
\item The Press-Schechter halo population is dependent on the minimum halo mass, i.e. the smallest halo which will house a galaxy. This is discussed in detail in Section~\ref{subsubsec:GAMADiscussion}. 
\item For this calculation we have assumed a uniform distribution of lens redshifts, which is not the case in real data. If the total lens mass distribution is specified however, any inhomogeneity in the galaxy distribution will average out over a large sample of source galaxies. Thus when comparing the results of our theoretical calculation to lensing probabilities in real data, it is acceptable to use the redshifts of the source galaxies in the sample and treat them as homogenous. Such a comparison is discussed in greater detail in Section~\ref{subsubsec:GAMADiscussion}. 
\end{itemize}
In spite of minor limitations, the results shown in Figure~\ref{fig:Probability_of_lensing_afo_zs} are strong motivation for a more through calculation and estimation of the probability of detecting weak lensing, since we are interested primarily in relatively low redshift weak lensing, where the DSM technique will be useful. 

\section{The lensing frequency algorithm}
\label{sec:Lensingfreqalg}

We now describe an algorithm for estimating the lensing signal per galaxy in a catalogue. We can use this algorithm to estimate the probability of any given object in the catalogue being lensed by some value, and to identify potentially suitable targets for lensing studies.\\

\noindent There are three factors which will determine the strength of the shear signal a lens imposes on a source: the mass of the lens, the angular separation of the lens and source, and the redshifts of the lens and source. Different combinations of these can lead to systems in which the lens appears large, and almost obscuring the source, but has a shear of the same magnitude as well separated systems with a small but dense lens. The simplest approach to estimating the strength of the shear present in a potential lens-source system is to simply inspect their separations and redshifts. Clearly this is prone to misidentifications, since systems may be falsely rejected using this method if the lens is particularly dense, and systems may be falsely selected if the lens appears large and close to a background source, but is of a very low density. \\

\noindent In the lensing frequency algorithm the shear probability is estimated utilising information about the stellar mass of the lens, and the redshifts and angular separations of each lens-source pair.  There are a number of steps involved in this process, beginning with the stellar masses of the galaxies in the catalogue, assuming a Singular Isothermal Sphere density distribution for each galaxy, estimating the halo circular velocities of each galaxy, and ultimately returning an estimate of the shear each object imposes on its nearest neighbours. These measurements can then be binned in shear to obtain an estimate of the probability distribution function for shear, and to flag particularly promising targets for follow up and direct measurement of the shear.  In this investigation we focus on galaxy-galaxy lensing only, however this process is equally applicable to galaxy-group lensing, provided good group masses are available. 

\subsection{Estimation of the circular velocity, $V_\mathrm{c}$}
\label{sec:Calc_of_V_circ}
The most uncertain step in the estimation of the estimated shears for each object is the calculation of the asymptotic circular velocity of the lens, $V_c$. This step is important as it takes us from the stellar mass to the total mass of each lens. Three approaches to this calculation were used:

\begin{enumerate}
\item Use a power law relation between halo mass and halo circular velocity obtained from the Bolshoi simulations \citep{2011ApJ...740..102K}. In this case it is necessary to compute the halo masses from the stellar masses, which is done using the relation derived by \citet{2010ApJ...710..903M}.
\item Assume the circular velocity at the outermost regions of the disk is approximately equal to the circular velocity in the halo. In this case, the baryonic Tully-Fisher relation \citep{2000ApJ...533L..99M} can be used to obtain $V_c$ from the known stellar masses of the objects.
\item Use an empirically derived relation between $V_\mathrm{c}$ and $\sigma_0$ from \citet{2007ApJ...655L..21C}. This relation has a significant scatter.
\end{enumerate}

There are significant uncertainties in all of the methods discussed here, and the discrepancies in the results obtained are at times significant (in particular for very high and very low mass objects). However, there is no single best method, and each method approaches the problem from a different starting point. These three methods may bracket the reality, and so by utilising all three methods simultaneously a good representation of possible values is obtained. Therefore the shear resulting from \itshape all three\upshape~methods are computed. \\

\noindent The steps taken by the algorithm are as follows:
\begin{enumerate}
\item The objects are sorted by redshift and the angular size distance for each object is computed by integrating over the Friedman equation:
\begin{equation}
D = \frac{c}{H_0} \int [\Omega_\mathrm{m} (1+z)^{3} + \Omega_\Lambda ]^{-1} \mathrm{d}z.
\end{equation} 

\item Beginning with the lowest redshift galaxy, assume it is a lens, and perform the following steps:

\begin{enumerate}
\item Compute the angular separation between the object and its neighbours out to a specified projected radius $\theta_\mathrm{max}$, excluding those at a lower redshift:
\begin{equation}
\theta_\mathrm{sep} = \sqrt{\Delta[R\cos D]^2 + \Delta [D]^2},
\end{equation}
where $\Delta[R\cos D] = (R_1 - R_2)\cos (D_1 + D_2)$,  $\Delta D = D_1-D_2$ and $R_i$, $D_i$ are an objects right ascension and declination\footnote{The purpose of only including galaxies with $\theta_\mathrm{sep} < \theta_\mathrm{max}$ in this step is to improve computation time, otherwise every object would be compared to every other higher redshift object. Background objects with a large separation from the source are likely to have negligible shears, and so can be safely excluded.}
. 
\label{it:N_neighbours}
\item Sort the neighbours by angular distance from the object;
\item Compute the halo mass from the stellar mass \citep{2010ApJ...710..903M}
\begin{equation}
\hspace*{-4pt}  
\frac{m(M_h)}{M_h} = 2 \left(\frac{m}{M_h}\right)_0\left[ \left(\frac{M_h}{M_1}\right)^{-\phi} + \left(\frac{M_h}{M_1}\right)^\eta \right]^{-1}\label{eq:M*_Mh_rel}
\end{equation}
where $M_h$ is the halo mass, m is the stellar mass, $(m/M_h)_{0}$ is a normalisation, $M_1$ is a characteristic mass where $m(M_h)/M_h$ is equal to $(m/M_h)_{0}$, and $\phi$ and $\eta$ are two slopes which indicate the behaviour of the relation at the low and high mass ends. The values for the free parameters used in this work are the best-fit values from \citet{2010ApJ...710..903M}: 
\begin{equation}
\begin{split}
\log{(M_1)} & = 11.884^{+0.030}_{-0.023} \\
\left(\frac{m}{M_h}\right)_0 & = 0.02820^{+0.00061}_{-0.00053}  \\
\phi & = 1.057^{+0.054}_{-0.046}  \\
\eta & = 0.556^{+0.010}_{-0.004} ; \\
\end{split}
\end{equation}
\item Compute the circular velocity three different ways: 
\begin{enumerate}
\item Using a power law relation derived from the Bolshoi simulation \citep{2011ApJ...740..102K}, which uses a $\Lambda$CDM cosmology:
\begin{equation}
V_\mathrm{c} = 2.8 \times 10^{-2} M_\mathrm{vir} ^{0.316}\label{eq:LCDM_meth},
\end{equation}
where $M_\mathrm{vir}$ is the virial mass and it is assumed that $M_\mathrm{vir} = M_h$.
\item From the stellar mass using the baryonic Tully-Fisher relation \citep{2000ApJ...533L..99M}
\begin{equation}
V_\mathrm{c} = \left(\frac{M_*}{{A}}\right)^{0.25}\label{eq:bTF_meth},
\end{equation}
where $A = 26.25\,h_{70}^{-2}\,M_{\odot}\,$km$^{-4}\,$s$^{4}$.
\item Using an empirically derived relation between $V_\mathrm{c}$ and $\sigma_0$ \citep{2007ApJ...655L..21C}:
\begin{equation}
V_\mathrm{c} = \sqrt{2}\sigma_0\label{eq:sig_0_meth}.
\end{equation}
\end{enumerate}
These will henceforth be referred to as the $\Lambda$CDM, bTF and $\sigma_0$ methods respectively.
\item The Einstein radius of the object is computed for each of the three $V_\mathrm{c}$ values (where it is assumed that due to the virial theorem one can write $V_c \sim \sigma_v$), and the shear for each $V_\mathrm{c}$ value is computed;
\end{enumerate}

\item Move to the next highest redshift galaxy, assume it is a lens, and repeat the above steps. 
\item Continue iterating, increasing in redshift until the second-highest redshift has been reached. 
\end{enumerate}

\section{Application to the GAMA survey data release 2 catalogue}
\label{sec:ApplicationtotheGAMAsurveycatalogue}


The target selection procedure outlined in Section~\ref{sec:Lensingfreqalg} has been applied to the Galaxy and Mass Assembly Phase 1 Survey (GAMA I) Data Release 2 (DR2) catalogue (\citeauthor{2011MNRAS.413..971D} \citeyear{2011MNRAS.413..971D}, Liske et al. in prep). The GAMA Survey is part of a larger project aiming to exploit the latest generation of ground-based and space-borne survey facilities to study cosmology and galaxy formation and evolution \citep{2009A&G....50e..12D}. Phase I of the GAMA Survey is a magnitude limited spectroscopic survey measuring galaxy spectra and redshifts in three equatorial regions centred at $9^{h}$, $12^{h}$ and $14.5^{h}$ (called G09, G12 and G15 respectively), each with an area of $12\times 4$ deg$^2$  \citep{2010MNRAS.404...86B} . The fields were observed to a limiting r-band apparent magnitude of $r_\mathrm{app} < 19.4$, $r_\mathrm{app} < 19.8$ and $r_\mathrm{app} < 19.4$ mag respectively. The target galaxies are distributed over a redshift range $0 < z \lesssim 0.5$ with a median redshift of $z \simeq 0.17$. The GAMA DR2 catalogue contains all GAMA I main survey objects down to $r<19.0$ mag (for G09 and G12) and $r<19.4$ mag (for G15) including spectral redshifts \citep{2014MNRAS.441.2440B}. The catalogue contains a total of 72,225 objects, of which 71,599 have derived stellar masses \citep{2011MNRAS.418.1587T}. GAMA survey data is available on the GAMA website\footnote{http://www.gama-survey.org}. \\

\noindent The lensing frequency algorithm has been applied to the GAMA I DR2 catalogue in two ways:
\begin{enumerate}
\item As the entire catalogue (with some minor cuts detailed below) to investigate the probability of any object being sheared by at least some value, and to identify conceivably suitable targets for potential follow up observation and shear measurement with DSM;
\item As a smaller sample to investigate the possibility of measuring the scatter in the $M_*-M_h$ relation, and to determine what sample size of galaxies is required to make this measurement.
\end{enumerate}

\noindent These two analyses are described in detail in the following subsections.

\subsection{Identifying weak lensing candidates in the GAMA survey Catalogue}
\label{subsec:targets_in_GAMA}
In this section we look at the entire GAMA 1 DR2 catalogue (with minor cuts described in the next paragraph), to investigate the probability of detecting shear of at least $\gamma$ for any given object in the catalogue, and identify candidate targets for follow up observation and shear measurement with DSM. 

\noindent After removing objects with undefined or uncertain redshift, and selecting for stellar mass in the range $8<\log_{10}(M_*/M_\odot)<12$, a dataset of 69,434 objects was obtained. The data was passed to the lensing frequency algorithm in two ways; 

\begin{figure}
\begin{center}
\includegraphics[width=8.5cm]{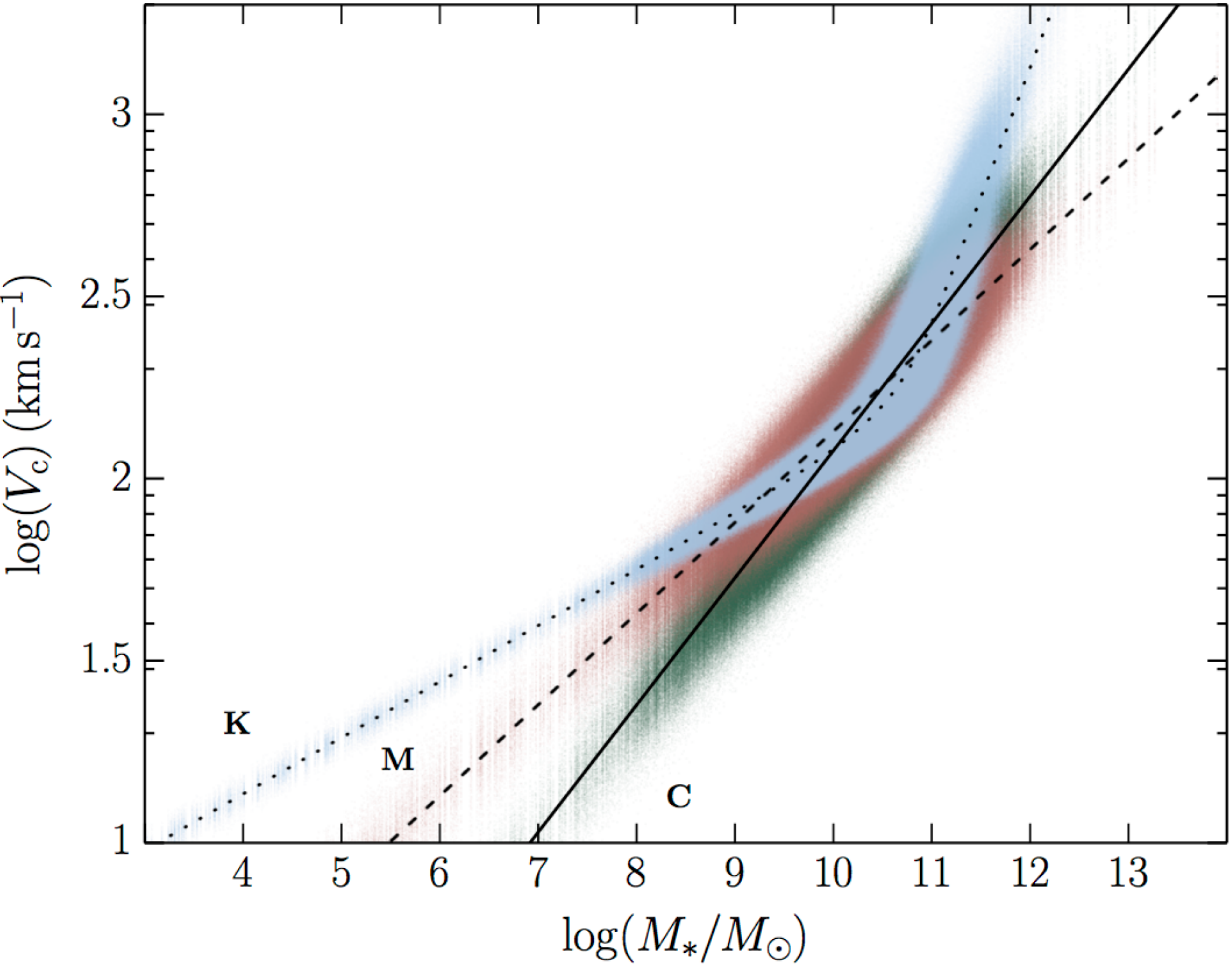}
\caption{ The $M_*-V_\mathrm{c}$ relation for three methods of calculation. `K' denotes the $\Lambda$CDM method \citep{2011ApJ...740..102K}, `M' the bTF method \citep{2000ApJ...533L..99M} and `C' the $\sigma_0$ method \citep{2007ApJ...655L..21C}. The dotted, dashed and solid lines show the relation with no scatter introduced. The blue, maroon and green points show a synthetic dataset obtained by making 100 realisations of the original dataset, and introducing a scatter of  $\sigma_{\Lambda\mathrm{CDM}} = 0.15$ dex, $\sigma_\mathrm{bTF} = 0.14\,h_{70}^{-2}$ dex, and $\sigma_{\sigma_0} = 0.08$ dex for the $\Lambda$CDM \citep{2010ApJ...710..903M}, bTF \citep{2000ApJ...533L..99M} and $\sigma_0$ methods respectively. The three methods agree well in the range $8<\log_{10}(M_*/M_\odot) <12$ (to within $\sim 15\%$), where most galaxies are situated. }\label{fig:Vcirc_vs_M_star}
\end{center}
\end{figure}

\begin{figure}
\begin{center}
\includegraphics[width=8.5cm]{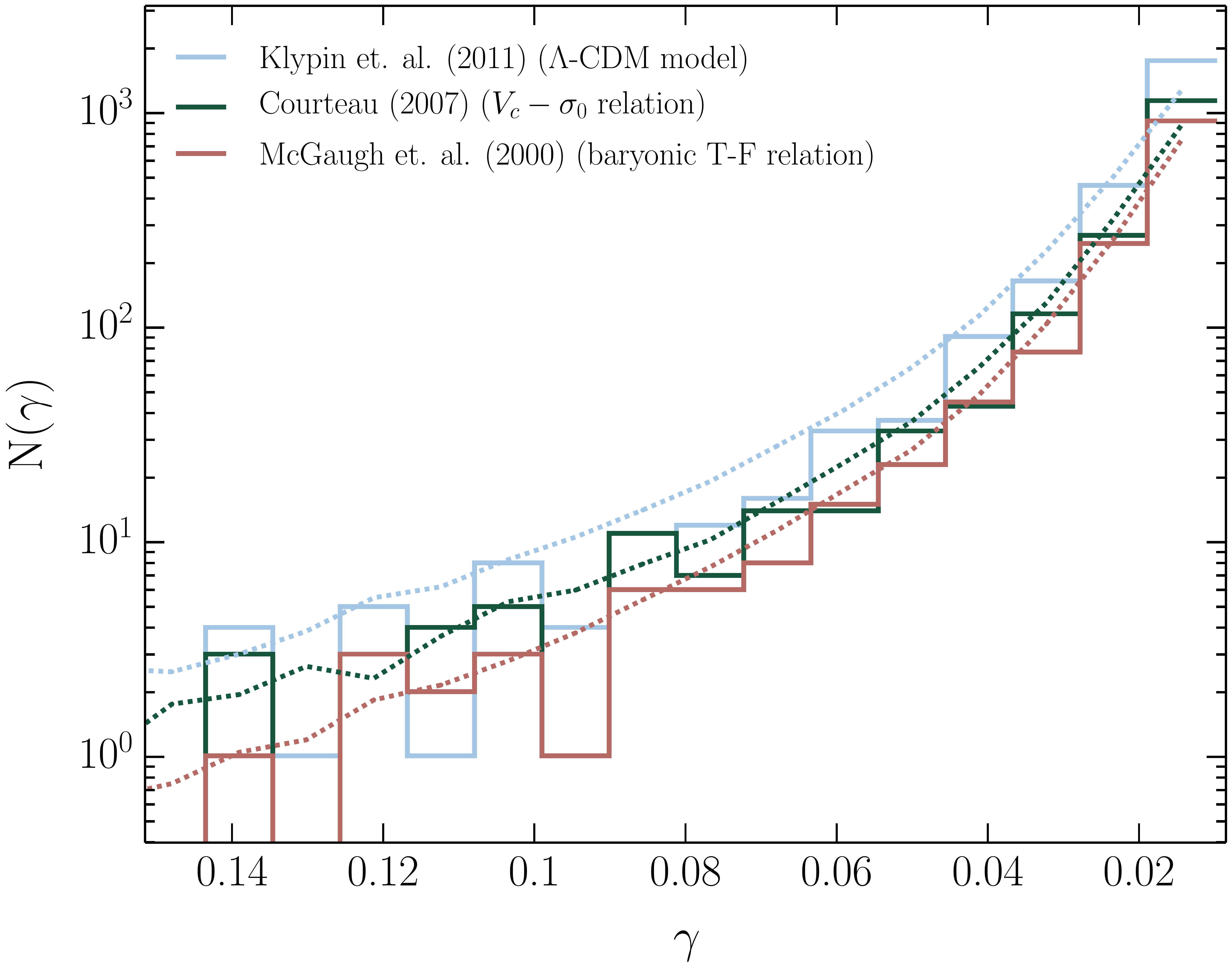}
\caption{ Histogram of estimated shears for galaxies in the GAMA survey. The solid, coarsely-binned lines correspond to the shears present in the original GAMA DR2 sample. The dotted lines correspond to the shears present in a synthetic dataset, obtained by producing 100 relisations of the original dataset, and introducing a scatter in the $M_*-V_\mathrm{c}$ relation. For every galaxy in the GAMA survey, there are 100 galaxies in the synthetic dataset. Therefore for ease of comparison of the histograms of the synthetic and real datasets, the number of sheared objects in the synthetic dataset have been divided by 100. One can see that the three methods agree well \itshape within~\upshape each of the real and synthetic datasets, with the synthetic datasets sitting slightly above the real datasets. }
\label{fig:raw_LCDMdist_bTFdist_sigma_0_together}
\end{center}
\end{figure}

\begin{itemize}
\item As the original set of galaxies (i.e. 69,434 objects). The shears present in the dataset were calculated via the three halo circular velocity estimates ($\Lambda$CDM, bTF and $\sigma_0$). The three $M_*-V_\mathrm{c}$ relations are shown in Figure~\ref{fig:Vcirc_vs_M_star} as dotted ($\Lambda$CDM), dashed (bTF) and solid ($\sigma_0$) lines. A histogram of shears present in the GAMA DR2 sample is shown in Figure~\ref{fig:raw_LCDMdist_bTFdist_sigma_0_together} as solid lines. The probability of a given galaxy being sheared by a particular value is obtained by normalising by the total number of galaxies in the sample.  
\item As a larger (synthetic) population of galaxies, to account for the scatter in the $M_*-V_\mathrm{c}$ relation. The synthetic population is produced by making 100 realisations of each input galaxy, with a scatter introduced to the realisations (giving a total dataset of 7,029,800 objects). The scatter in the population arises from the intrinsic scatter in each $M_*-V_\mathrm{c}$ relation and so is different for each of the three relations used. A scatter of $\sigma_{\Lambda\mathrm{CDM}} = 0.15$ dex was introduced into Equation~(\ref{eq:M*_Mh_rel}) for the $\Lambda$CDM method \citep{2010ApJ...710..903M}; a scatter of $\sigma_\mathrm{bTF} = 0.14\,h_{70}^{-2}$ dex was introduced into Equation~(\ref{eq:bTF_meth}) for the bTF method \citep{2000ApJ...533L..99M}; and a scatter of $\sigma_{\sigma_0} = 0.08$ dex was introduced into Equation~(\ref{eq:sig_0_meth}) for the $\sigma_0$ method. The $\sigma_0$ values and associated scatter used in the $\sigma_0$ method were obtained by fitting for the stellar masses using velocity dispersions obtained from the Sloan Digital Sky Survey Data Release 10 (SDSSDR10; \citeauthor{2014ApJS..211...17A} \citeyear{2014ApJS..211...17A}). The $M_*-V_\mathrm{c}$ relation of the synthetic population of galaxies is shown in Figure~\ref{fig:Vcirc_vs_M_star} as blue ($\Lambda$CDM), maroon (bTF) and green ($\sigma_0)$ points. One can see that in the range $8<\log_{10}(M_*/M_\odot) <12$, where most galaxies are situated, the three methods agree well. The shears present in the synthetic dataset were computed for every point, and the resulting number of sheared objects, normalised by 100, are presented in Figure~\ref{fig:raw_LCDMdist_bTFdist_sigma_0_together}, as dotted lines. The synthetic data is normalised by 100 to allow for better comparison to the real data, since for every galaxy in the real dataset there are 100 galaxies in the synthetic dataset. One can see that the three methods agree well  \itshape within~\upshape within each of the real and synthetic datasets, with the synthetic datasets sitting slightly above the real datasets. The implications of this upward shift are discussed in Section~\ref{subsec:Scatter_in_M*_Mh_rel}. The probability of a given galaxy being sheared by a particular value is obtained by normalising by the total number of galaxies in the sample.
\end{itemize}


\subsubsection{Discussion}
\label{subsubsec:GAMADiscussion}

\noindent The number of galaxies with shears above the cutoff value of 0.02 varied between the three methods of calculating $V_\mathrm{c}$, with the $\Lambda$CDM method giving the largest estimate of measurable shears in the data, and the bTF method giving the smallest estimate. The probability of a shear of at least $\gamma = 0.02$ was $\mathrm{P}(\geq\gamma) = 0.018, 0.005$ and $0.007$ for the $\Lambda$CDM, bTF and $\sigma_0$ methods respectively. The bTF method was chosen for selecting candidate targets because it gives the smallest shear probability, and so will be the least likely to overestimate the shears present in a sample. Using the bTF method, the number of objects in GAMA with an estimated shear of $\gamma\geq 0.02$ is $393$. The bTF method always estimates a smaller shear than the $\Lambda$CDM and $\sigma_0$ methods for any given object, so any of the $393$ objects selected by the bTF method would also have been selected by the other two methods (although of course the converse is not true).\\ 

\begin{figure}
\begin{center}
\includegraphics[width=8.5cm]{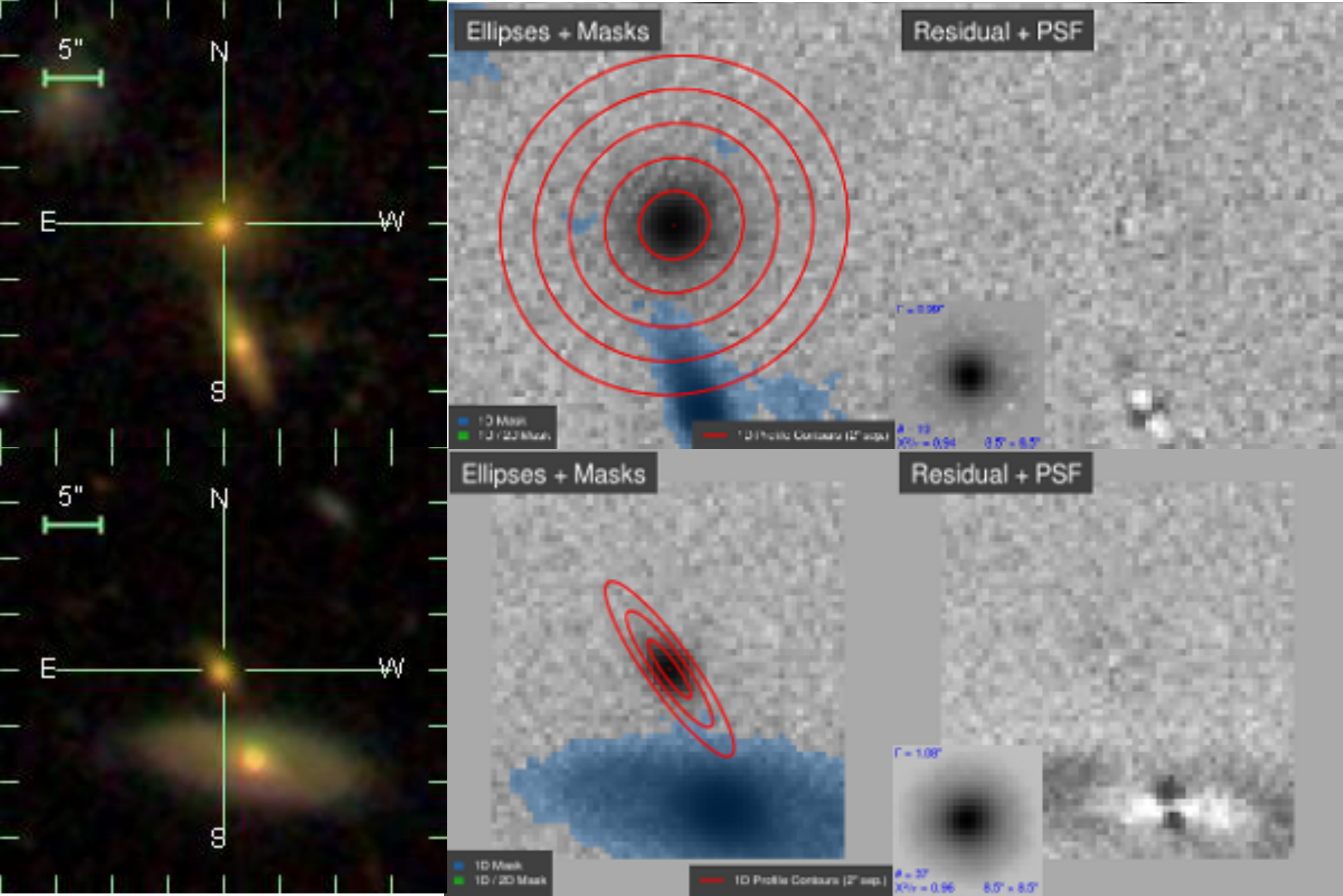}
\caption{ Two example galaxy pairs from the GAMA DR2 Sample identified with the target selection algorithm. In both cases the galaxy in the crosshairs is the source galaxy (i.e. the galaxy being lensed). The left-hand images show thumbnails of the galaxy pairs from the SDSS DR10 Finding Chart Tool, while the middle and right-hand images show the J-band images from the UKIDSS survey, with the residuals from 2D S\'{e}rsic fits, taken from GAMA's online Single Object Viewer tool. The top pair is at RA $= 213.705$ deg, DEC $= 1.623$ deg, and has an estimated shear of $\gamma = 0.023$. The lens and source redshifts are $z = 0.128$ and $z = 0.186$ respectively. The bottom pair is at RA $= 213.705$ deg, DEC $= 1.623$ deg, and has an estimated shear of $\gamma = 0.053$. The lens and source redshifts are $z = 0.088$ and $z = 0.190$ respectively. }
\label{fig:twopairs}
\end{center}
\end{figure}

\noindent These objects were extracted from the sample, and matched to objects in the SDSS DR10 catalogue. The lens-source pairs in the sample were ranked by each of the authors by eye, based on their morphology, surface brightness, inclination angle, separation and environment. The motivation for selecting on this criterion stems from the requirements of the DSM algorithm; that the source galaxy be undisturbed and stably-rotating, neither face-on nor edge-on and bright enough for observation with an IFU. For a more detailed discussion of the requirements of the DSM algorithm, see \citet{2015MNRAS.451.2161D}. Two highly-ranked example lens-source pairs are shown in Figure~\ref{fig:twopairs}.\\
  
\begin{figure}
\begin{center}
\includegraphics[width=8.5cm]{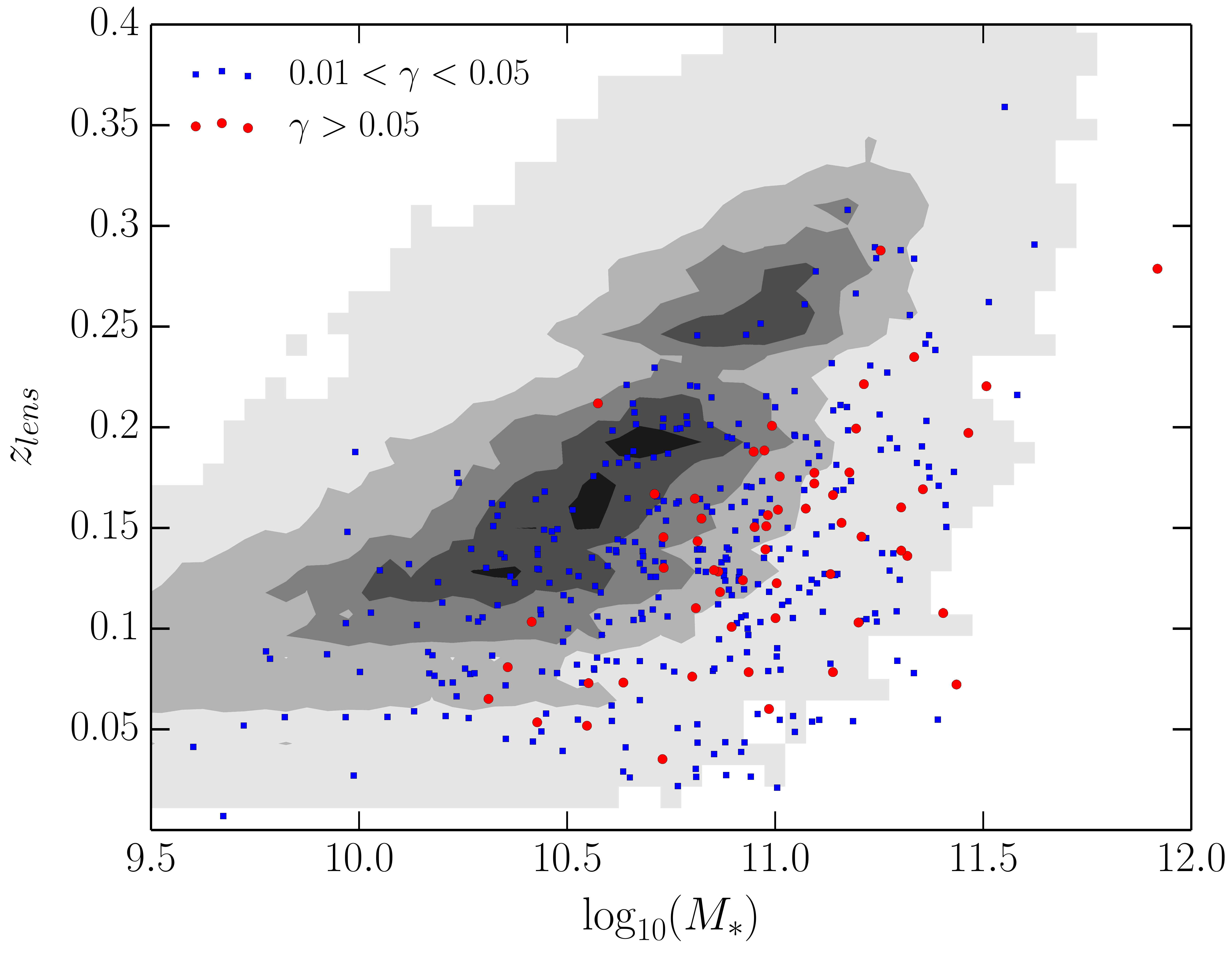}
\caption{Lens stellar mass as a function of lens redshift for the GAMA DR2 sample, showing the distribution of shears with these parameters. Grey contours show the density of all galaxies in the sample. Blue squares show galaxies with a shear in the range $0.01<\gamma<0.05$, and red circles show those with a shear in the range $\gamma>0.5$. The shears were estimated using the bTF method.} 
\label{fig:lens_mstar_v_z_w_gamma}
\end{center}
\end{figure}

\noindent Figure~\ref{fig:lens_mstar_v_z_w_gamma} shows the distribution of galaxies with significant estimated shear, with lens stellar mass plotted as a function of lens redshift. As might be expected, the selection favours more massive lensing galaxies, with significant numbers of candidates at all redshifts. The quality of the shear measurements however depends on the characteristics of the source galaxy. \\

\begin{figure}
\begin{center}
\includegraphics[width=8.5cm]{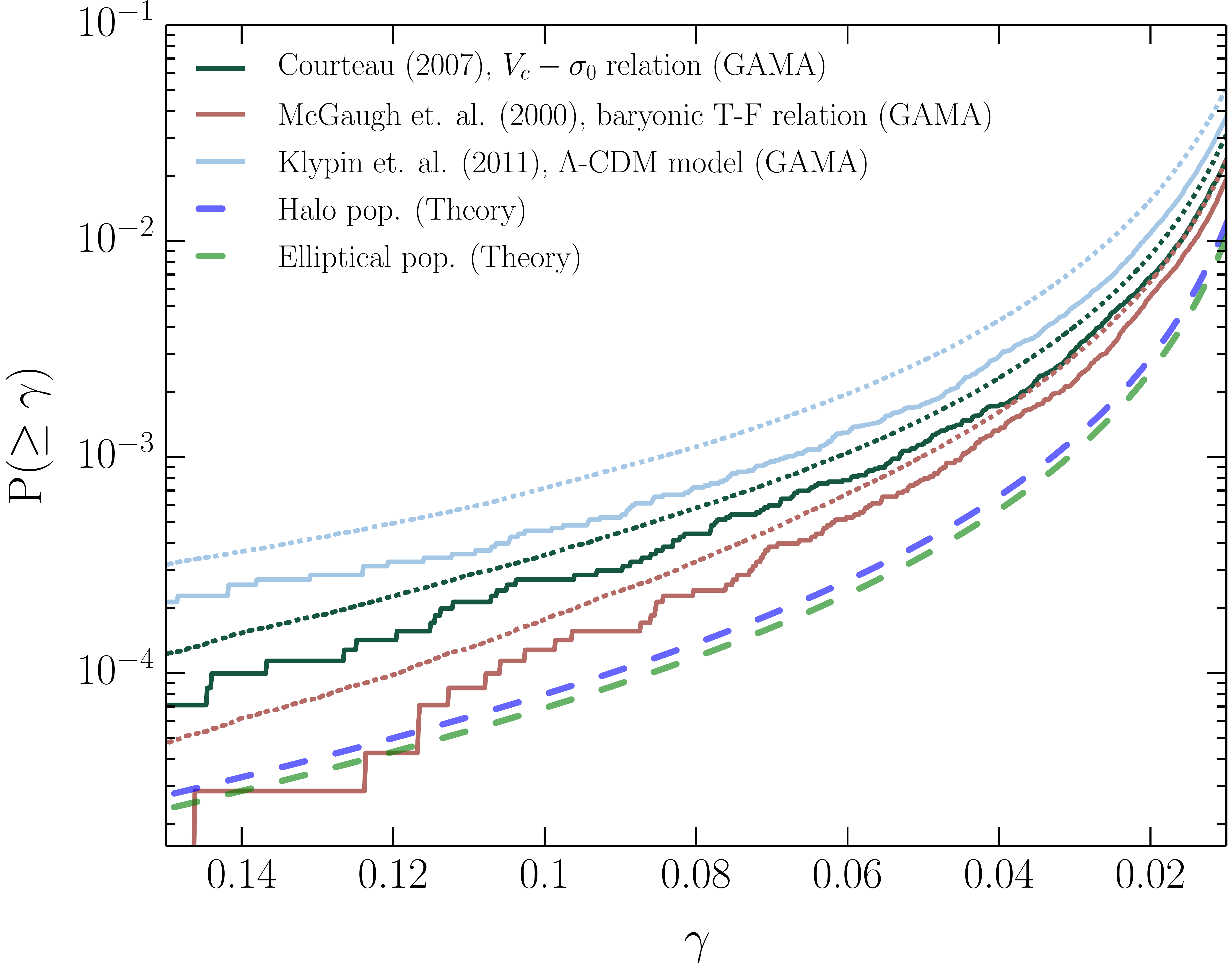}
\caption{ Estimated probability of measuring a shear of \itshape{at least}\upshape~$\gamma$ for galaxies in the GAMA survey. The dotted lines are the probabilities with scatter introduced into the $M_*-V_\mathrm{c}$ relation (i.e. the synthetic dataset), and the solid lines are the probabilities with no scatter introduced. Again one can see that the synthetic datasets sit higher than the real datasets. The dashed lines are included for comparison purposes, and correspond to the probabilities as a function of $\gamma$ derived from the theoretical calculation of probability in Section~\ref{sec:LensingProb}, using the redshifts of the sheared objects in the GAMA survey. }
\label{fig:cumPDF_LCDMdist_bTFdist_sigma_0_together_AT_LEAST_x}
\end{center}	
\end{figure}

\noindent The normalised cumulative sum was taken of the data shown in Figure~\ref{fig:raw_LCDMdist_bTFdist_sigma_0_together}, the results of which are shown in Figure~\ref{fig:cumPDF_LCDMdist_bTFdist_sigma_0_together_AT_LEAST_x}, giving the probability of measuring a shear of \itshape at least~\upshape $\gamma_\mathrm{lim}$. One can again see the upward shift in the distributions for synthetic data, relative to those for the real data. Also shown are two dashed lines corresponding to our theoretical estimate of the probability of lensing from Section~\ref{sec:LensingProb}. The probability of at least a given shear being observed in a galaxy increases as the minimum shear decreases. Since the lensing cross-section of each galaxy goes as $\gamma_\mathrm{lim}^{-2}$, this is not surprising, as smaller minimum shears rapidly increase the fraction of the sky in which measurable lensing will occur. The method which gives the lowest overall probability is the bTF method. Since we expect to be able to measure shears as small as $\gamma = 0.02$ with DSM, we can expect to have \itshape{at least}~\upshape a one in two hundred chance of measuring shear in a randomly chosen galaxy (from the bTF method), and up to a roughly one in fifty chance (from the $\Lambda$CDM method). \\

\noindent It is obvious that the $\Lambda$CDM method results in much higher probabilities than the bTF and $\sigma_0$ methods. It is interesting to note that the $\Lambda$CDM method is the one which uses only simulation and theoretically-derived relations to obtain values of $V_\mathrm{c}$ from $M_*$. In contrast, the bTF and $\sigma_0$ methods utilise empirically derived relationships between $V_\mathrm{c}$ and $M_*$. One can easily identify the origin of the higher numbers of lensed galaxies from the $\Lambda$CDM method; the break in the $M_* - V_\mathrm{c}$ relation for this method. While giving good agreement in the intermediate mass range, this results in a much larger corresponding halo circular velocity. These higher $V_\mathrm{c}$ galaxies will have a much larger lensing cross section, resulting in a larger number of lensed objects. Hence, while there are relatively few galaxies with $\log_{10}(M_*/M_\odot)>12$, they contribute strongly to the total lensing probability. This raises the question: how does this over-abundance of higher-$V_\mathrm{c}$ galaxies affect the results of $\Lambda$CDM-based cosmological simulations? Further consideration of this topic is beyond the scope of this paper, and is left to future work.\\

\begin{figure}
\begin{center}
\includegraphics[width=8.5cm]{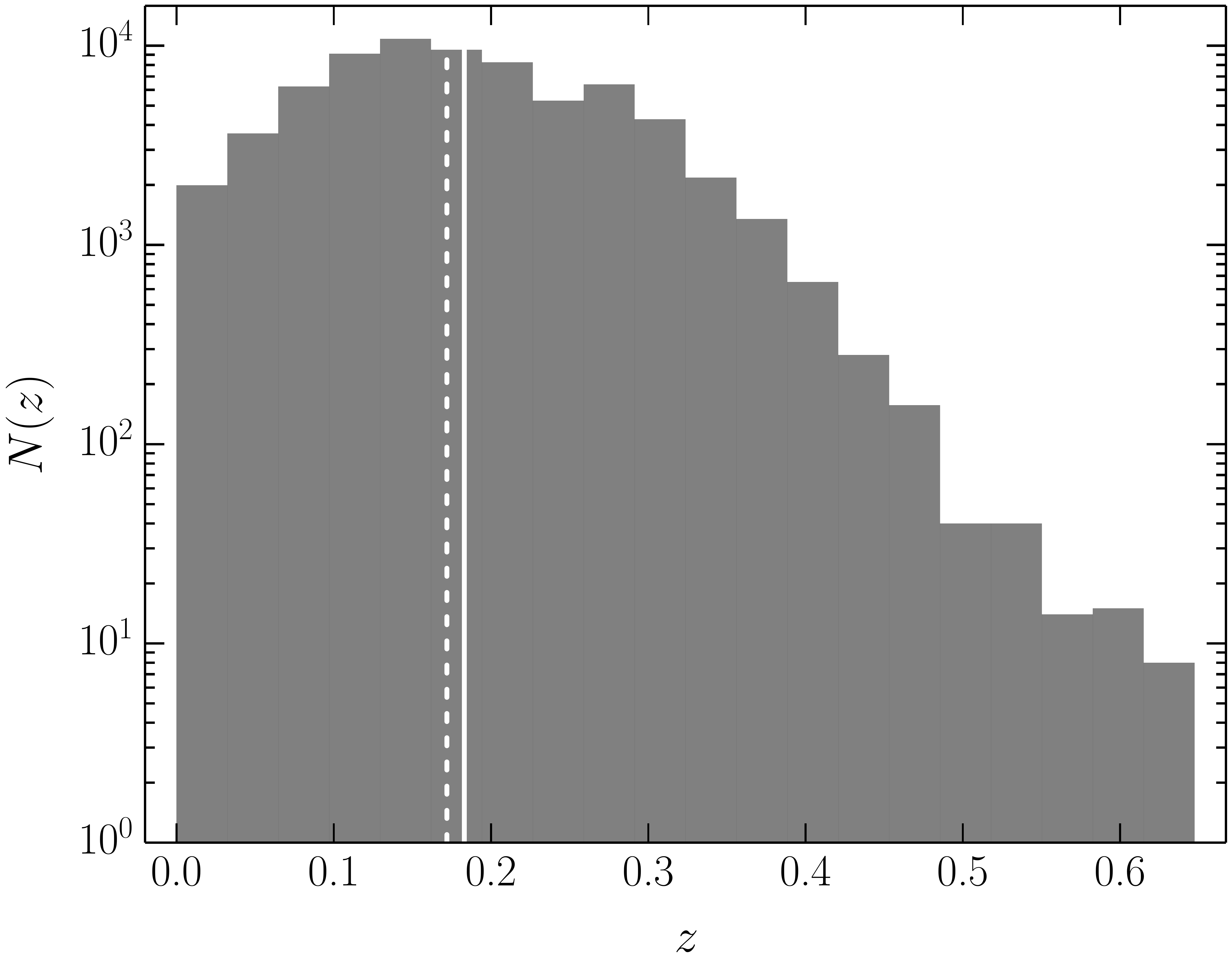}
\caption{ Distribution of redshifts in the GAMA sample, after a redshift quality cut has been made. The mean and median of this distribution are $0.18$ (solid white line) and $0.17$ (dashed white line) respectively, and the range of the distribution is $0\lesssim z\lesssim 0.65$. }
\label{fig:hist_of_z}
\end{center}
\end{figure}


\noindent As mentioned already, in Figure~\ref{fig:cumPDF_LCDMdist_bTFdist_sigma_0_together_AT_LEAST_x} we have included two dashed lines corresponding to our theoretical estimate of the probability of lensing from Section~\ref{sec:LensingProb}. The probability calculated in Section~\ref{sec:LensingProb} is the probability that a source at redshift $z_s$ is lensed by at least some value $\gamma_\mathrm{lim}$. The probability obtained from the lensing frequency algorithm however is the probability that a single galaxy from the GAMA DR2 sample, chosen at random, is lensed by at least $\gamma_\mathrm{lim}$. In order to compare the two, we need to calculate the theoretical probability that one galaxy, chosen at random from a GAMA-like population of galaxies, is lensed by at least $\gamma_\mathrm{lim}$. From Section~\ref{sec:LensingProb}, the probability that some object in the GAMA DR2 sample is lensed by at least $\gamma_\mathrm{lim}$ will be dependent on its redshift only (once a lens mass distribution has been assumed). Thus the probability that \itshape{any}~\upshape galaxy in the sample has been lensed by at least $\gamma_\mathrm{lim}$ will be the sum of the individual probabilities of each one having been lensed. Then the probability that one galaxy chosen at random has been lensed by at least $\gamma_\mathrm{lim}$ will be the probability of any having been lensed, divided by the total number of galaxies in the sample. Figure~\ref{fig:hist_of_z} shows the histogram of the redshifts of the GAMA DR2 sample after a redshift quality cut has been made. While the distribution of redshifts in the GAMA sample is not uniform, it is sufficiently smoothly varying that the assumption made in Section~\ref{sec:LensingProb} (that the lenses are uniformly distributed in redshift) will suffice. Thus we have arrived at a method by which we can compare the results of Section~\ref{sec:LensingProb} and Section~\ref{subsec:targets_in_GAMA}.\\

\noindent From Section~\ref{sec:LensingProb}, the theoretical probability of measuring a shear of at least $\gamma = 0.02$ in any one galaxy in the GAMA survey is $\mathrm{P}(\geq\gamma) \simeq 0.002$ and $0.003$ for the halos housing elliptical galaxies and Press-Schechter halo populations respectively. Comparing to the probabilities obtained from the GAMA DR2 sample ($\mathrm{P}(\geq\gamma) = 0.018, 0.005$ and $0.007$ for the $\Lambda$CDM, bTF and $\sigma_0$ methods respectively), we see the difference is a factor of $\sim 6$ for the $\Lambda$CDM method, and a factor of $\sim 2$ for the bTF and $\sigma_0$ methods. This is an acceptable level of agreement, given the assumptions made in the calculations in Section~\ref{sec:Lensingfreqalg}. The level of agreement between the two approaches can be seen in Figure~\ref{fig:cumPDF_LCDMdist_bTFdist_sigma_0_together_AT_LEAST_x}. \\

\begin{figure}
\begin{center}
\includegraphics[width=8.5cm]{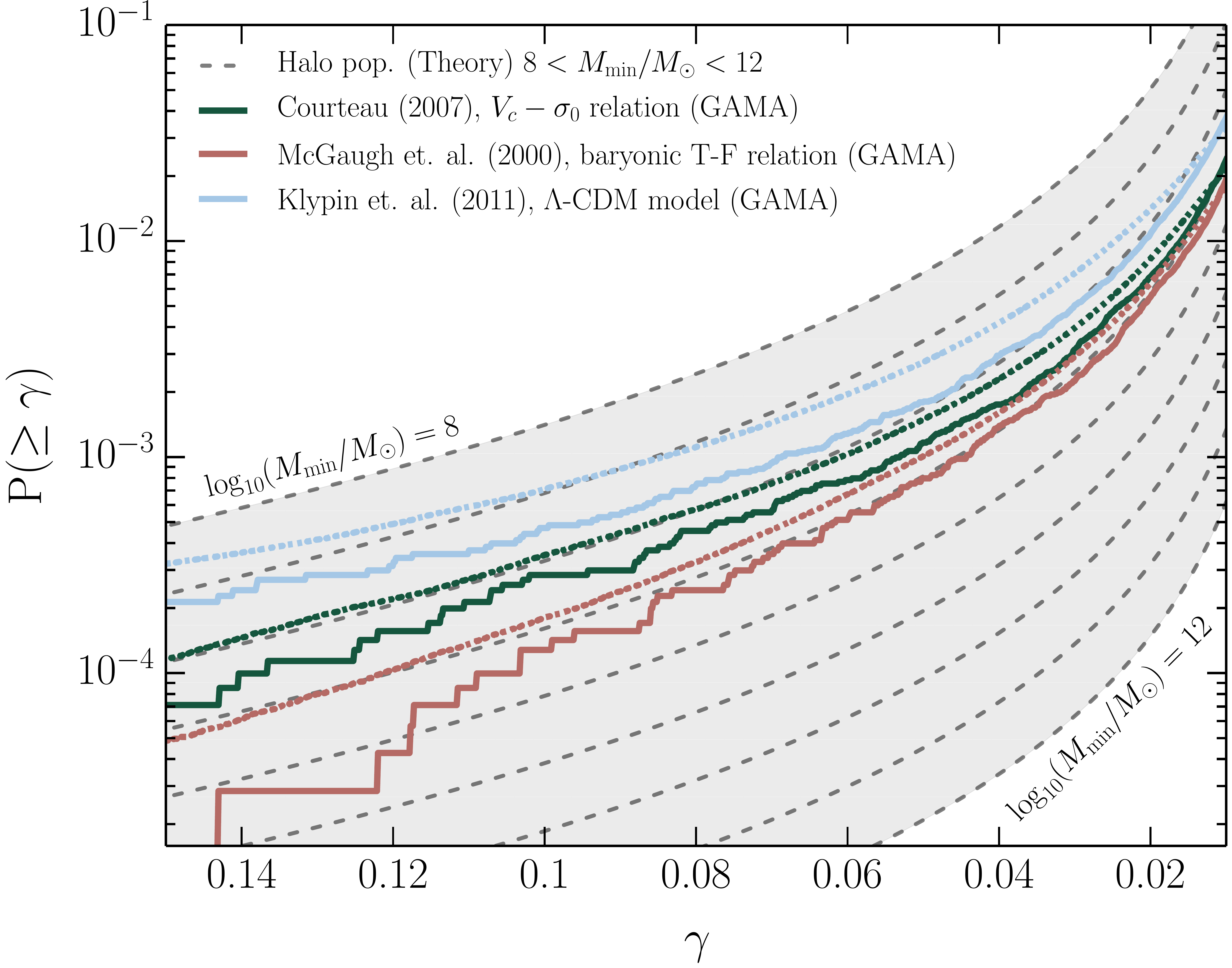}
\caption{Comparison of the results from the lensing frequency algorithm with the theoretically derived Press-Schechter probabilities, with varying $M_\mathrm{min}$. The highest theoretically-derived probability corresponds to a minimum halo mass of $\log_{10}(M_\mathrm{min}/M_\odot) = 8$, while the lowest corresponds to $\log_{10}(M_\mathrm{min}/M_\odot) = 12$. The results of the lensing frequency algorithm are plotted as in Figure~\ref{fig:cumPDF_LCDMdist_bTFdist_sigma_0_together_AT_LEAST_x}. The bTF and $\sigma_0$ methods best agree with a Press-Schechter halo population with a minimum halo mass in the range $9\lesssim \log_{10}(M_\mathrm{min}/M_\odot)\lesssim 10$. }
\label{fig:GAMA_cf_varying_M_star}
\end{center}	
\end{figure}

\noindent The theoretical estimate of $P(>\gamma_\mathrm{lim})$ drawn from the Press-Schechter halo population is dependent on the value of the minimum halo mass in which a galaxy will form. As can be seen in Figure~\ref{fig:Probability_of_lensing_afo_zs}, a smaller (larger) value of $M_\mathrm{min}$ will result in a higher (lower) probability of measuring a shear of at least $\gamma_\mathrm{lim}$ as a function of redshift. It can be seen in Figure~\ref{fig:GAMA_cf_varying_M_star} that adjusting $M_\mathrm{min}$ will alter the level of agreement between the theoretical estimate and the probabilities derived from the GAMA DR2 sample. Each dashed grey line represents a minimum halo mass incremented by $0.5\log_{10}(M_\odot)$. The line with the highest corresponding probability is that with $\log_{10}(M_\mathrm{min}/M_\odot)  = 8$,  and that with the lowest corresponding probability $\log_{10}(M_\mathrm{min}/M_\odot)  = 12$. The probabilities from the lensing frequency algorithm are as in Figure~\ref{fig:cumPDF_LCDMdist_bTFdist_sigma_0_together_AT_LEAST_x}. It is interesting to note that for the bTF and $\sigma_0$ methods, the best agreement with the theoretical approach is obtained for a minimum halo mass in the range $9\lesssim \log_{10}(M_\mathrm{min}/M_\odot)\lesssim 10$, whereas the currently accepted value is $\log_{10}(M_\mathrm{min}/M_\odot) = 10$. 

\subsection{Measuring the scatter in the $M_*-M_h$ relation}
\label{subsec:Scatter_in_M*_Mh_rel}

Since halos of  given mass can have different halo concentrations, spin parameters and merger histories, we expect them to house galaxies with a range of masses. This manifests as a scatter in the stellar mass to halo mass relation. The value of the scatter in the $M_*-M_h$ relation is not well constrained, as directly measuring the masses of galaxies \itshape and~\upshape their host halos is not trivial. Abundance matching techniques can be useful for describing the relationship between stellar mass and halo mass, however they cannot constrain the scatter in the $M_*-M_h$ relation \citep{2009ApJ...696..620C}. \citet{2012ApJ...744..159L} used traditional weak lensing techniques to constrain the $M_*-M_h$ relation, however as with abundance matching they are unable to place any meaningful constraints on the scatter in the relation. An alternative approach has been to utilise satellite galaxies to constrain the galaxy luminosity-halo mass relation. Early work involved stacking the central galaxies to obtain a statistical measure of the kinematics of the satellite galaxies \citep{1987Natur.325..779E,1993ApJ...405..464Z,1994ApJ...435..599Z,1997ApJ...478...39Z}. However, recent work has avoided the need for stacking, largely by utilising the larger datasets offered by the 2dF Galaxy Redshift Survey (2dFGRS; \citeauthor{2001MNRAS.328.1039C} \citeyear{2001MNRAS.328.1039C}) and SDSS \citep{2002ApJ...571L..85M,2003ApJ...593L...7B,2003ApJ...598..260P,2004MNRAS.352.1302V,2004bdmh.confE..41V,2005ApJ...635..982C,2007ApJ...654..153C,2009MNRAS.392..801M}. These techniques to not utilise weak lensing, and do not attempt to measure the scatter in the $M_*-M_h$ relation.\\

\noindent In this section we test whether DSM measurements of a population of galaxies could be used to measure the scatter in the $M_*-M_h$ relation. This is a novel measurement which is difficult with traditional weak lensing techniques, but is made possible with DSM because it can measure individual shears with far greater accuracy around individual galaxies. We then briefly describe how to fit for the scatter in the $M_*-M_h$ relation, and measure the scatter in a set of simulated shear datasets with a known scatter incorporated.\\

\noindent As was noted in Section~\ref{subsec:targets_in_GAMA}, and can be seen in Figures~\ref{fig:raw_LCDMdist_bTFdist_sigma_0_together} and \ref{fig:cumPDF_LCDMdist_bTFdist_sigma_0_together_AT_LEAST_x}, scatter in the $M_*-M_h$ relation results in a shift in the distribution of shears present in a population of galaxies towards larger shears. If the measurement error in the shear is sufficiently small, or the population of galaxies with known stellar masses is large enough, it is possible to measure the scatter in the $M_*-M_h$ relation by comparing the distribution of shears to those obtained from a $M_*-M_h$ relation with zero scatter.\\

\noindent To perform this measurement with DSM, velocity maps from from an intermediate redshift survey would be required. The velocity maps can be obtained from several components of the galaxy, such as HI in radio wavelengths, or Integral Field Unit (IFU) observations of stellar velocities and gas emission in optical wavelengths. IFU maps obtained from bright, H$\alpha$ emitting galaxies are the most practical and easily obtainable in the immediate term however, and so we will focus on these galaxies in this analysis.\\

\noindent We assume that the scatter in the $M_*-M_h$ relation is lognormal, which is the standard form assumed in the literature (e.g. \citeauthor{2010ApJ...717..379B} \citeyear{2010ApJ...717..379B}, \citeauthor{2010ApJ...710..903M} \citeyear{2010ApJ...710..903M}, \citeauthor{2013ApJ...770...57B} \citeyear{2013ApJ...770...57B}). The mean of a lognormal distribution is a function of the size of the scatter in the distribution, and is given by $\exp(\mu+\sigma^2/2)$, where $\sigma$ is the scatter in the distribution, and $\mu$ is the natural logarithm of the mean of the underlying normal distribution. It can be seen that the mean increases with increasing scatter. Thus if a lognormal scatter is introduced into the $M_*-M_h$ relation for a population of galaxies, the result is a larger number of galaxies with higher mass halos. It is this property of the lognormal scatter in the $M_*-M_h$ relation which leads to an increased probability of larger shears.\\

\noindent To compare the shear distributions obtained from populations with and without scatter in the $M_*-M_h$ relation, the lensing frequency algorithm was applied to a selection of mock catalogues generated from a subsample of the complete GAMA DR2 sample. This subsample was obtained by performing the same cuts on the GAMA DR2 as in the previous section, along with the following additional cuts: $0.1<z<0.15$, $r<17.5$, and keeping only galaxies with H-$\alpha$ emission. The resulting dataset contained 2,861 galaxies. When computing shears in the mock catalogues, a maximum lens-source separation of $D_\mathrm{max} = 0.2$ Mpc was used for step~(\ref{it:N_neighbours}) of the lensing frequency algorithm. \\

\noindent A successful measurement of the scatter in the $M_*-M_h$ relation will be limited by survey sample size and the size of the scatter. To investigate the range in which these parameters would enable a measurement of the scatter, a set of simulated datasets of varying size was created, with a range of scatters in the $M_*-M_h$ relation.\\

\noindent To produce the mock catalogues, larger populations of galaxies were generated from the subsample of GAMA DR2 galaxies by producing Monte-Carlo realisations of each galaxy. The resulting datasets contained $N$= 1,000, 15,000, 50,000 and 150,000 galaxies. The distribution of shears for a tight $M_*-M_h$ relation was computed for the simulated datasets. A scatter, $\sigma_{M_*-M_h}$, was then included in the $M_*-M_h$ relation and the shears were computed again. This process was repeated for a selection of values of $\sigma_{M_*-M_h}$ between $0.1$ and $0.5$. This range of scatter is chosen to bracket the current literature values (e.g.. \citeauthor{2009ApJ...693..830Y} \citeyear{2009ApJ...693..830Y}, \citeauthor{2009MNRAS.392..801M} \citeyear{2009MNRAS.392..801M}, \citeauthor{2010ApJ...717..379B} \citeyear{2010ApJ...717..379B}, \citeauthor{2010ApJ...710..903M} \citeyear{2010ApJ...710..903M}). The resulting `true' shears for each dataset were then `observed' with a range of measurement errors, so that the `observed' shear in the tight datasets contained a shear measurement error only, while the `observed' shear in the scattered datasets contained a shear measurement error, and additional scatter from the $M_*-M_h$ relation.\\

\begin{figure*}
\begin{center}
\includegraphics[width=17cm]{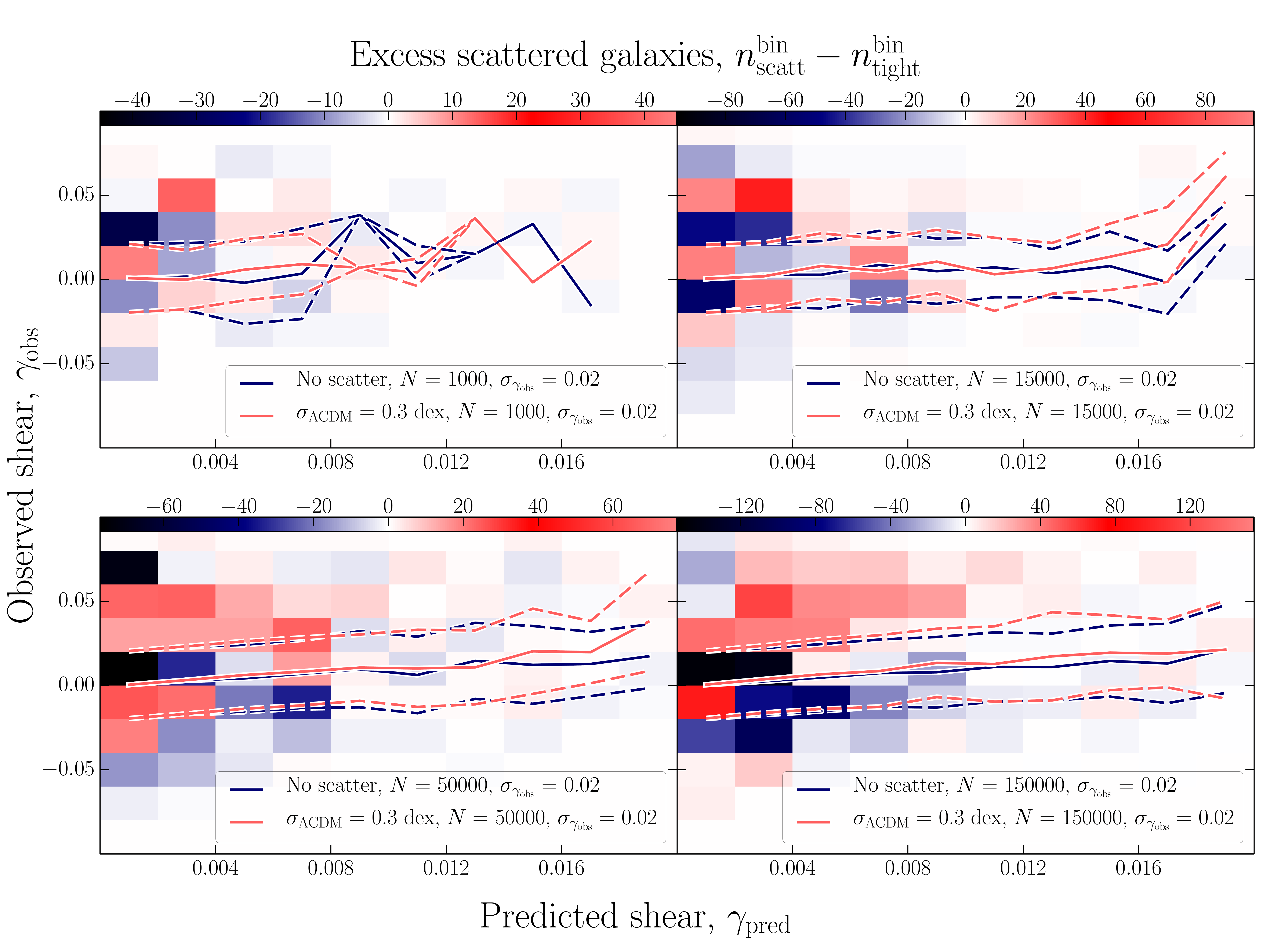}
\caption{Distributions of observed shears as a function of predicted shears with and without scatter in the $M_*-M_h$ relation, for $N =$ 1,000, 15,000, 50,000 and 150,000 respectively (clockwise from top left), with $\sigma_{\gamma_\mathrm{obs}}=0.02$ and $\sigma_{M_*-M_h} =0.3$ dex. The solid blue and red lines show the mean in the tight and scattered distributions respectively. The dashed blue and red lines show $\pm1\sigma$ from the mean in the tight and scattered distributions respectively. The shaded background (and colourbar at the top of each plot) shows the relative excess (red) or shortfall (blue) of galaxies with scatter relative to galaxies without scatter, $n^\mathrm{bin}_\mathrm{scatt}-n^\mathrm{bin}_\mathrm{tight}$, in 2D bins of size $(\Delta\gamma_\mathrm{obs}\times\Delta\gamma_\mathrm{pred}) = (0.02\times 0.002)$. }
\label{fig:fourfigs_w_n}
\end{center}
\end{figure*}

\begin{figure*}
\begin{center}
\includegraphics[width=17cm]{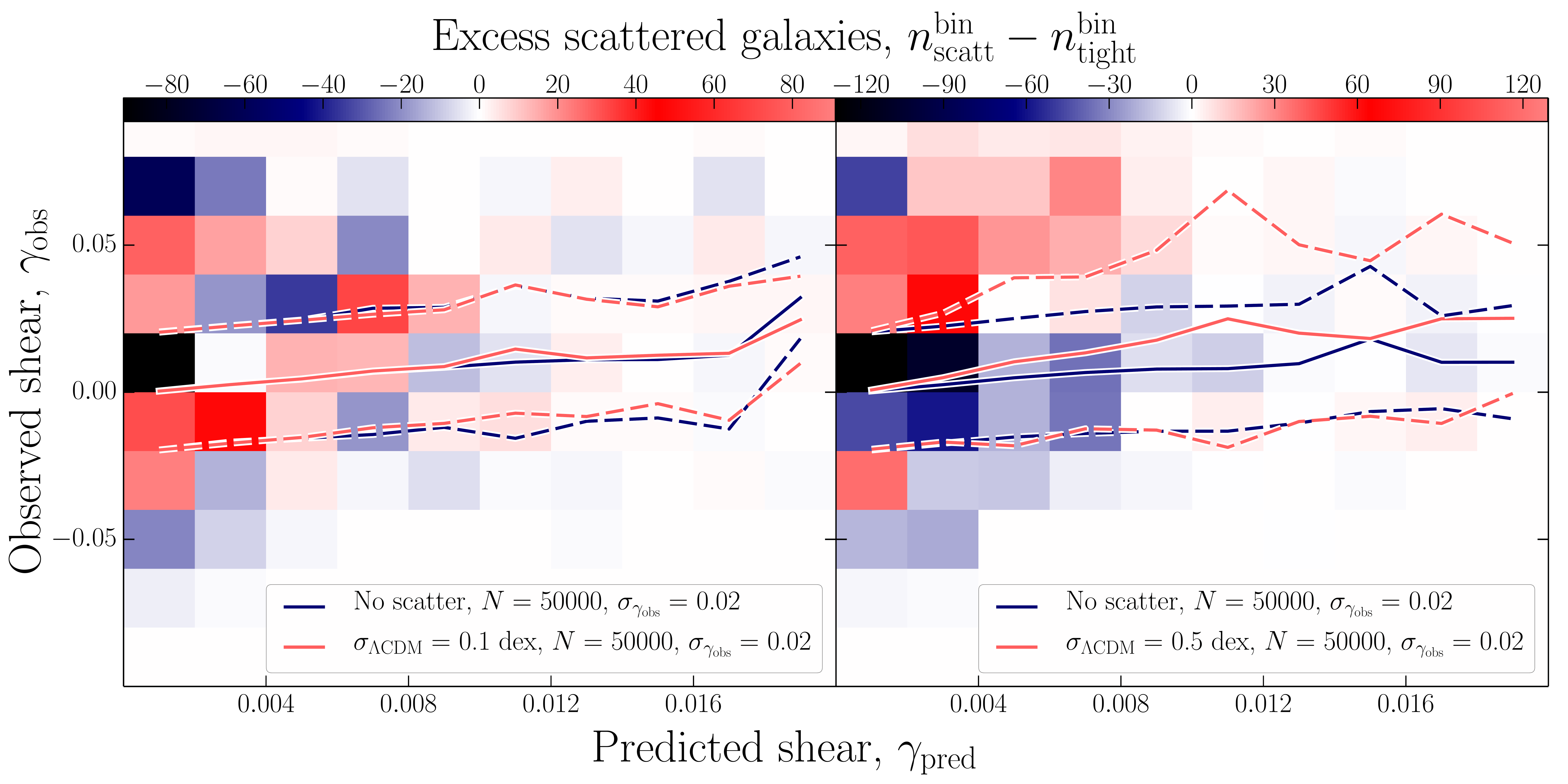}
\caption{Distributions of observed shears as a function of predicted shears with and without scatter in the $M_*-M_h$ relation, for $\sigma_{M_*-M_h} = 0.1$ dex (left plot), and 0.5 dex (right plot), with $\sigma_{\gamma_\mathrm{obs}}=0.02$ and $N$ = 15,000. All lines and shaded regions are as in Figure~\ref{fig:fourfigs_w_n}. }
\label{fig:twofigs_w_scatt}
\end{center}
\end{figure*}

\begin{figure*}
\begin{center}
\includegraphics[width=17cm]{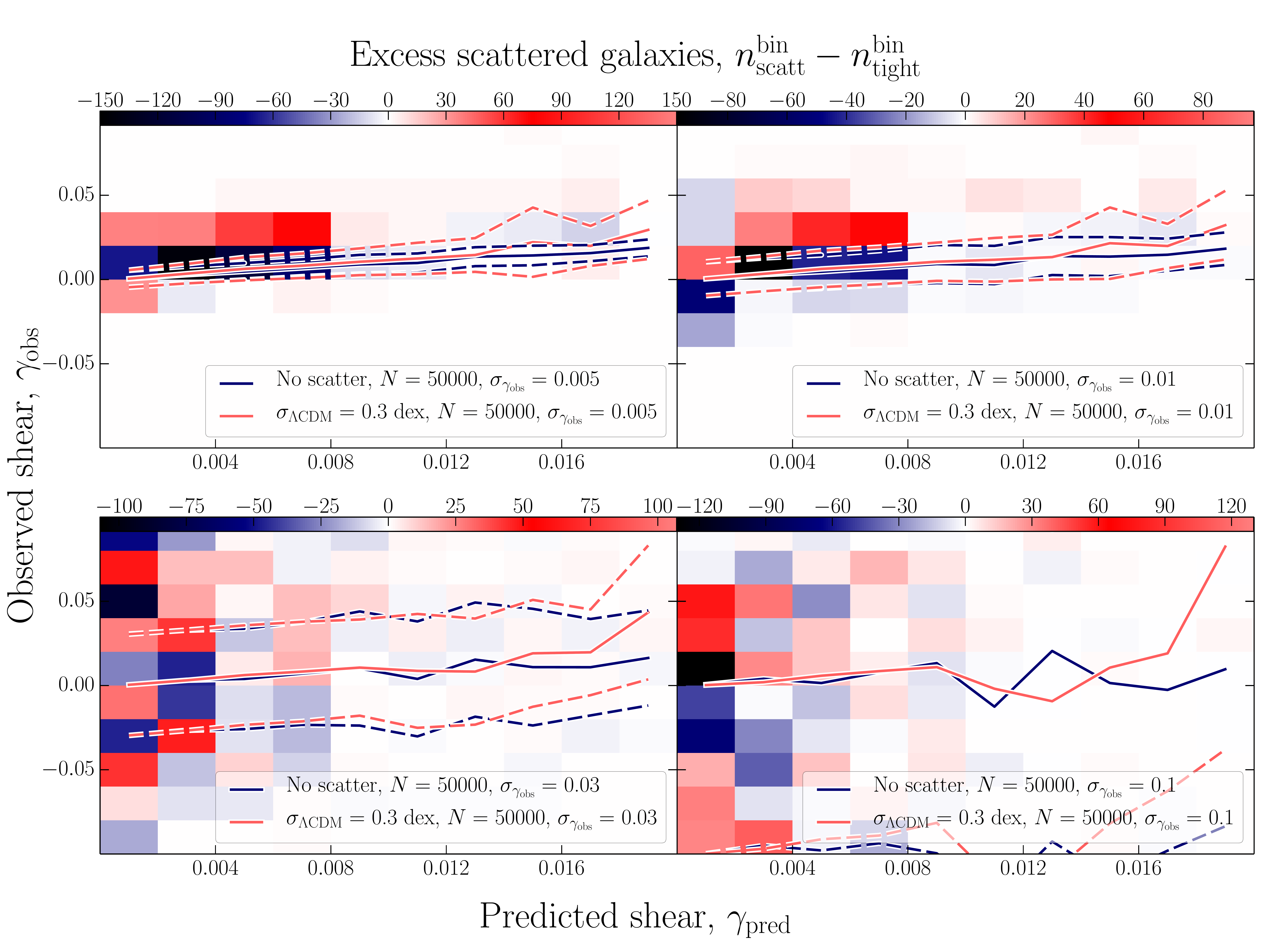}
\caption{Distributions of observed shears as a function of predicted shears with and without scatter in the $M_*-M_h$ relation, for $\sigma_{\gamma_\mathrm{obs}}$ = 0.005, 0.01, 0.03 and 0.1, with $\sigma_{M_*-M_h} = 0.3$ dex and $N$ = 15,000. All lines and shaded regions are as in Figure~\ref{fig:fourfigs_w_n}. }
\label{fig:fourfigs_w_obs}
\end{center}
\end{figure*}

\noindent The key questions are: 1) Can we identify the scatter from the $M_*-M_h$ relation over the scatter from measurement error? 2) How many objects do we need to do so? 3) What is the uncertainty on the measurement? To answer these questions, one first needs to establish that for a given number of galaxies, the distribution of shears arising from a tight $M_*-M_h$ relation can be distinguished from the distribution with scatter in the $M_*-M_h$ relation. The distributions of shears with and without scatter in the $M_*-M_h$ relation are shown in Figures~\ref{fig:fourfigs_w_n}-\ref{fig:fourfigs_w_obs}. Figure~\ref{fig:fourfigs_w_n} shows the observed shear as a function of the predicted shear for $N =$ 1,000, 15,000, 50,000 and 150,000, for a shear measurement error of $\sigma_{\gamma_\mathrm{obs}}=0.02$ and $M_*-M_h$ relation scatter of $\sigma_{M_*-M_h} =0.3$ dex. The solid blue and red lines show the mean in the tight and scattered distributions respectively, and the dashed blue and red lines show $\pm1\sigma$ from the mean in the tight and scattered distributions respectively. The shaded background shows the relative excess (red) or shortfall (blue) of galaxies with scatter relative to galaxies without scatter, in 2D bins of size $(\Delta\gamma_\mathrm{obs}\times\Delta\gamma_\mathrm{pred}) = (0.02\times 0.002)$. Figure~\ref{fig:twofigs_w_scatt} shows the observed shear as a function of the predicted shear for $N =$ 50,000, shear measurement error of $\sigma_{\gamma_\mathrm{obs}}=0.02$, and scatters of $\sigma_{M_*-M_h} =0.1$ and $0.5$ dex. Lines and shading are as in Figure~\ref{fig:fourfigs_w_n}. Figure~\ref{fig:fourfigs_w_obs} shows the observed shear for $\sigma_{\gamma_\mathrm{obs}}$ = 0.005, 0.01, 0.03, and 0.1 for $N =$ 50,000 and $\sigma_{M_*-M_h}= 0.3$dex. Lines and shading are as in Figure~\ref{fig:fourfigs_w_n}. \\

\noindent The distributions of observed shears with and without scatter in the $M_*-M_h$ relation were compared using the Two Sample Kolmogorov-Smirnov (2SKS) Test. For it to be possible for the scatter in the $M_*-M_h$ relation to be measurable from comparing the distribution of observed to predicted shears with and without scatter, we require the distributions to fail the hypothesis that they are drawn from the same distribution, under the 2SKS Test. That is, we require the p-value of the test to be small. The resulting 2SKS scores and p-values (denoted $S_\mathrm{2SKS}$ and $P_\mathrm{2SKS}$) for the combinations of $n$ and $\sigma_{M_*-M_h}$ considered are presented in Table~\ref{table:2SKSreults}.\\

\begin{table*}
\caption{The Two Sample Kolmogorov-Smirnov Test score, $S_\mathrm{2SKS}$, and p-value, $P_\mathrm{2SKS}$, for the scattered and tight datasets, for each combination of $N$, $\sigma_{\gamma_\mathrm{obs}}$ and $\sigma_{M_*-M_h}$ investigated. $n$ is the number of simulated datapoints for each real galaxy,  $N$ is the total number of simulated datapoints, and $\sigma_{M_*-M_h}$ is the scatter in the $M_*-M_h$ relation.
}
\centering
\label{table:2SKSreults}
\begin{tabular}{c c c c c }
\hline
\\[-0.5ex]
$\mathbf{N}$ & $\boldsymbol{\sigma_{\gamma_\mathrm{obs}}}$ & $\boldsymbol{\sigma_{M_*-M_h}}$ & $\boldsymbol{S_\mathrm{2SKS}}$ & $\boldsymbol{P_\mathrm{2SKS}}$ \\ [0.5ex]
\hline
\\[-0.5ex]
1,000      & 0.02 & $ 0.3 $ & $2.5\times 10^{-2}$ & 0.91  \\
15,000   & 0.02 & $ 0.3 $ & $8.7\times 10^{-3}$  & 0.61 \\ 
50,000   & 0.02 & $ 0.3 $ & $1.1\times 10^{-2}$  & $4.4\times 10^{-3}$  \\ 
150,000 & 0.02 & $ 0.3 $ & $7.9\times 10^{-3}$ & $1.9\times 10^{-3}$ \\[1ex]
\hline 
\\[-0.5ex]
50,000 & 0.005 & $ 0.3 $ & $1.6\times 10^{-2}$ & $6.2\times 10^{-6}$   \\
50,000 & 0.01 & $ 0.3 $   & $1.4\times 10^{-2}$ & $1.4\times 10^{-4}$ \\ 
50,000 & 0.03 & $ 0.3 $   & $1.0\times 10^{-2}$ & $9.5\times 10^{-2}$  \\ 
50,000 & 0.1 & $ 0.3 $     & $8.7\times 10^{-3}$ & $4.7\times 10^{-2}$ \\[1ex]
\hline 
\\[-0.5ex]
50,000 & 0.02 & $ 0.1 $  & $7.6\times 10^{-3}$ & 0.11 \\
50,000 & 0.02 & $ 0.5 $  & $1.9\times 10^{-2}$ & $1.6\times 10^{-8}$  \\[1ex]
\hline 
\end{tabular}
\end{table*}

\begin{figure}
\begin{center}
\includegraphics[width=8.5cm]{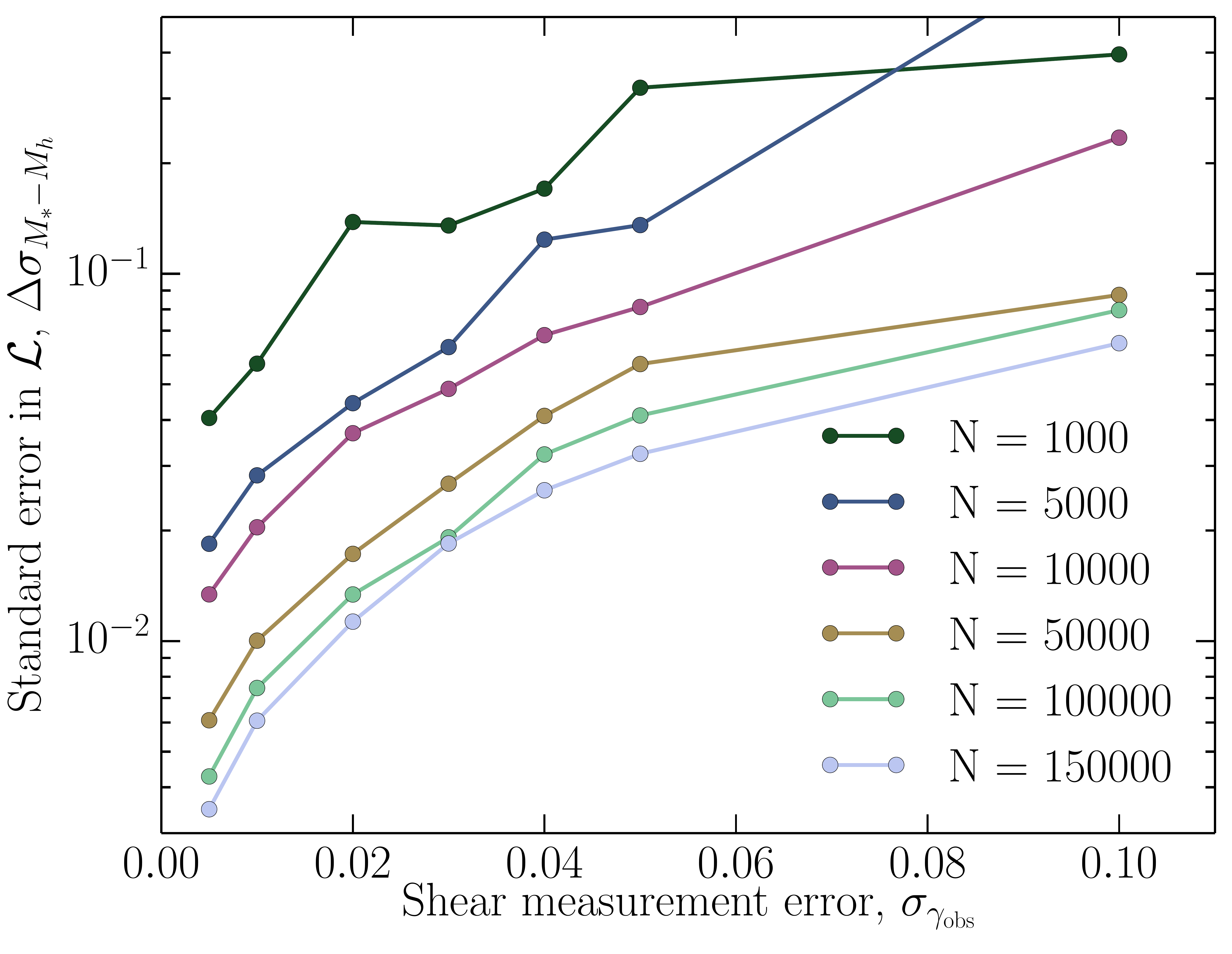}
\caption{The standard error in the maximum value of $\log[\mathcal{L}(\sigma_{M_*-M_h})]$ as a function of the shear measurement error, $\sigma_{\gamma_\mathrm{obs}}$ for values of $N$ in the range 1,000$<N<$150,000. }
\label{fig:stand_err_plot}
\end{center}
\end{figure}

\noindent The next step is to fit for the scatter in $M_*-M_h$ relation in simulated datasets. To do this, we assume that the observed shears are distributed according to both the scatter in the $M_*-M_h$ relation, and an observation error, so that the likelihood of an observed shear given the true underlying shear is given by
\begin{equation}
\begin{split}
L(\gamma_\mathrm{obs}&|\gamma_\mathrm{true}, \sigma_{M_*-M_h}, \sigma_{\gamma_\mathrm{obs}})\\
&= \int N(\tau - \gamma_\mathrm{obs},0,\sigma_{\gamma_\mathrm{obs}}) \\
&\times M(\tau,\gamma_\mathrm{true},\sigma_{M_*-M_h})\mathrm{d}\tau,
\end{split}
\end{equation}
where $N$ is a Gaussian distribution representing the measurement error in the DSM method, and $M$ is a lognormal distribution representing the scatter in the $M_*-M_h$ relation. $N$ is given by
\begin{equation}
N(x,\mu,\sigma) = \frac{1}{\sqrt{2\pi}\sigma}\exp{\left[-\frac{1}{2}\left( \frac{x-\mu}{\sigma} \right)^2\right]}
\end{equation}
where $\mu$ is the mean of the distribution and $\sigma$ is the standard deviation. The lognormal, $M$, is a distribution whose logarithm is normally distributed. It is given by
\begin{equation}
M(x,\mu,\sigma) = \frac{1}{\sqrt{2\pi}\sigma x}\exp{\left[-\frac{1}{2}\left( \frac{\log (x/\mu)}{\sigma} \right)^2\right]}\label{eq:lognormal}
\end{equation}
where $\mu$ and $\sigma$ are the mean and standard deviation of the underlying normal distribution. The total likelihood of a given value of $\sigma_{M_*-M_h}$ for fixed $\sigma_{\gamma_\mathrm{obs}}$ is then given by
\begin{equation}
\mathcal{L}(\sigma_{M_*-M_h}) = \sum_i L(\gamma_{\mathrm{obs},i}|\gamma_{\mathrm{true},i},  \sigma_{M_*-M_h} ,\sigma_{\gamma_\mathrm{obs}}).
\end{equation}
We have fitted for the scatter in the $M_*-M_h$ relation for an assumed true scatter of $\sigma_{M_*-M_h}=0.3$ by maximising $\log[\mathcal{L}(\sigma_{M_*-M_h})]$ for a range of values of $N$ and $\sigma_{\gamma_\mathrm{obs}}$ to investigate the behaviour of the uncertainty in the fit with these parameters. A Fisher Matrix analysis was used to estimate the standard error in the maximum likelihood for each $\sigma_{\gamma_\mathrm{obs}}$ and $N$. Figure~\ref{fig:stand_err_plot} shows the standard error, $\Delta\sigma_{M_*-M_h}$, as a function of these parameters. We find that for an assumed scatter of $\sigma_{M_*-M_h}=0.3$, to obtain a robust fit with a measurement error of $\sigma_{\gamma_\mathrm{obs}}=0.02$ a dataset of $N\sim$ 50,000 DSM measurements is required, a result which is consistent with the results of the 2SKS test.\\

\subsubsection{Discussion}
\label{subsubsec:Scatt_in_M*_Mh_rel}
\noindent For fixed $\sigma_{\gamma_\mathrm{obs}}$ and $\sigma_{M_*-M_h}$, the p-value for $N =$ 15,000 is too large to rule out the hypothesis that the two samples of data come from the same distribution, while the p-value for $N =$ 50,000 is sufficient to rule out this hypothesis, and the p-value for $N =$ 150,000 can easily do so. Similarly, for fixed $N$ and $\sigma_{M_*-M_h}$, the p-value for $\sigma_{\gamma_\mathrm{obs}}$ = 0.03 is too large to confidently rule out the hypothesis that the two samples of data come from the same distribution, while the p-value for $\sigma_{\gamma_\mathrm{obs}}$ = 0.02 is sufficient to rule out this hypothesis, and the p-value for $\sigma_{\gamma_\mathrm{obs}}$ = 0.01 can easily do so. Not surprisingly, a larger scatter in the $M_*-M_h$ relation results in a smaller p-value, since a larger scatter directly increases the difference between the datasets with and without scatter in them. From the above considerations we conclude that in order to measure a scatter of $\sigma_{M_*-M_h} = 0.3$, a sample of $\gtrsim$ 50,000 galaxies with a shear measurement error of $\sigma_{\gamma_\mathrm{obs}} \lesssim$ 0.02 would be required. For larger values of the scatter, larger values of $\sigma_{\gamma_\mathrm{obs}}$ and smaller values of $N$ would be sufficient. \\

\noindent This result is confirmed in our fits for a scatter in the $M_*-M_h$ relation of $0.3$ for a range of values of $N$ and $\sigma_{\gamma_\mathrm{obs}}$, presented in Figure~\ref{fig:stand_err_plot}. For a true $M_*-M_h$ scatter of $0.3$, a shear measurement error of $\sigma_{\gamma_\mathrm{obs}} = 0.02$, and a sample of $N=$50,000 shear measurements, we recover a fitted scatter of 0.308$\pm$0.02.\\

\noindent DSM is expected to achieve measurement errors of $\sigma_{\gamma_\mathrm{obs}} \sim$ 0.02, and so shear measurement error is not seen to be a limiting factor in measuring scatter in the $M_*-M_h$ relation. While a sample of $\gtrsim 50,000$ galaxies with spatial and spectral resolution does not yet exist, there are several surveys beginning in the near to intermediate future which will provide datasets of a sufficient size to perform this experiment, for example surveys with the Hector instrument \citep{2012SPIE.8446E..53L} on the Anglo-Australian Telescope (AAT), and The Hobby-Eberly Telescope Dark Energy Experiment (HETDEX; \citeauthor{2008ASPC..399..115H}, \citeyear{2008ASPC..399..115H}), or surveys on the Square Kilometre Array (SKA). We conclude that while it would not be possible to utilise DSM to measure the scatter in the $M_*-M_h$ relation with existing IFU survey data, it will be possible with data from upcoming surveys.

\section{Conclusions}
\label{sec:Conclusions}
We have made an analytical estimate of the frequency of a source being weakly lensed given a uniformly distributed population of lenses, following \citet{2000MNRAS.319..860M}. We have adapted their work for the weak lensing case, in which we consider the probability of the source being lensed by \itshape{at least}\upshape~some limiting value $\gamma_\mathrm{lim}$. The results of this analysis suggest the probability of detecting weak lensing greater than a limiting value of $\gamma_\mathrm{lim} = 0.02$ in a realistically observable redshift range ($z\lesssim 1$) is non-negligible. Given this, we have created a lensing frequency algorithm which searches an input dataset for all lens-source pairs with an estimated shear greater than a limiting value of  $\gamma_\mathrm{lim}$. Our algorithm has been applied to a dataset extracted from the GAMA survey catalogue, and the number of objects with an estimated shear of at least $\gamma = 0.02$ in the sample was found to be $\sim 393$. These targets can be matched to objects in the SDSS DR10 Catalogue, and a subsample of good targets can be chosen from this selection for follow up observations. \\

\noindent A scatter in the $M_*-M_h$ relation results in a shift towards higher measured shears for a given population of galaxies. Given this, we have investigated the feasibility of measuring the scatter in the $M_*-M_h$ relation using shear statistics. We find that for a given shear measurement error, our ability to differentiate between a distribution of shears from a `tight' $M_*-M_h$ relation, and one with scatter, is dependent on the size of the scatter, the number of objects in the sample, and the shear measurement error. For a scatter of $0.3$ dex in the $M_*-M_h$ relation, we find that a sample size of $\sim$50,000 galaxies would be needed to measure the scatter, for a measurement error on the shear of $0.02$ (a value consistent with the shear measurement accuracy achievable with DSM). We attempt to fit for the scatter in the $M_*-M_h$ relation for a set of simulated datasets. The result of this is shown in Figure~\ref{fig:stand_err_plot}. For a true scatter of $0.3$, a shear measurement error of $0.02$ and $\sim$50,000 shear measurements, we recover a scatter of 0.308$\pm$0.02. It should be noted that the technique we have demonstrated here is based on a relatively untested algorithm, however it is adequate as an illustrative example and our results are promising for future measurements. While there are no existing IFU survey catalogues of a sufficient size to apply this technique, there are several surveys beginning in the near to intermediate future which will provide datasets of a sufficient size to perform this experiment, for example HETDEX which aims to observe $\sim 10^6$ galaxies, or surveys on the SKA, such as the `billion galaxy survey' which aims to observe $\sim 10^9$ galaxies.

\begin{acknowledgements}
We thank the referee for their constructive feedback. Their input into this publication has greatly improved it. This research was conducted as part of the Australian Research Council Centre for Excellence for All-Sky Astrophysics (CAASTRO), through project number CE110001020. We acknowledge financial support from The University of Melbourne, and the Australian Astronomical Observatory (AAO). Special thanks goes to the members of the GAMA survey team for their advice, input and helpful discussions. GAMA is a joint European-Australasian project based around a spectroscopic campaign using the Anglo-Australian Telescope. The GAMA input catalogue is based on data taken from the Sloan Digital Sky Survey and the UKIRT Infrared Deep Sky Survey. Complementary imaging of the GAMA regions is being obtained by a number of independent survey programs including GALEX MIS, VST KiDS, VISTA VIKING, WISE, Herschel-ATLAS, GMRT and ASKAP providing UV to radio coverage. GAMA is funded by the STFC (UK), the ARC (Australia), the AAO, and the participating institutions. The GAMA website is \url{http://www.gama-survey.org/}. Funding for SDSS-III has been provided by the Alfred P. Sloan Foundation, the Participating Institutions, the National Science Foundation, and the U.S. Department of Energy Office of Science. The SDSS-III web site is \url{http://www.sdss3.org/}.
\end{acknowledgements}

\bibliographystyle{apj}
\bibliography{bibliography}

\label{lastpage}
\end{document}

%% file: aasmacros.tex
\providecommand{\apjl}{ApJ}                
\providecommand{\apjs}{ApJS}               
\providecommand{\ao}{Appl.~Opt.}           
\providecommand{\aap}{A\&A}                

